%% file: main_v2.tex
\documentclass[a4paper,11pt]{article}
\pdfoutput=1 

\usepackage{jheppub} 

\usepackage[T1]{fontenc} 

\usepackage{amssymb}
\usepackage{amsmath}
\allowdisplaybreaks

\usepackage{graphicx}
\usepackage{subcaption}

\usepackage{xltabular}
\usepackage{changepage}
\newcolumntype{C}{>{$}c<{$}}
\usepackage{booktabs}

\usepackage{multirow}

\usepackage{diagbox}

\usepackage{float}

\usepackage{mathtools}

\usepackage{bbding}
\usepackage{pifont}
\newcommand{\cmark}{\ding{51}}%
\newcommand{\xmark}{\ding{55}}%

\usepackage{tikz}
\usetikzlibrary{arrows.meta,decorations.markings,calc,positioning}
\usepackage{csquotes}

\usepackage[inline]{enumitem}

\newcommand{\disp}{\tilde{\mathbb{F}}}
\newcommand{\Fpri}{\hat{\mathbb{F}}}
\newcommand{\Fdesc}{\check{\mathbb{F}}}

\newcommand{\Fstar}{{\mathbb{F}}^{*}}

\newcommand{\fstar}{{{f}}^{*}}
\newcommand{\ftilde}[1]{{\tilde{f}}^{*_{#1}}}

\title{\boldmath 
A Rosetta Stone for Wilson Line Defects
}

\author{Julius Julius$\,{}^{\Phi_{||}}$}
\author{and Nika Sergeevna Sokolova$\,{}^{y_m^2,\,[y_m^2]_{2}}$}

\affiliation{
${}^{\Phi_{||}}$ Laboratoire de Physique de l'\'Ecole Normale Sup\'erieure, Universit{\'e} Paris Sciences et Lettres, Centre National de la Recherche Scientifique, Sorbonne Universit{\'e}, Universit{\'e} Paris Cit{\'e}, 24 rue Lhomond, 75005 Paris, France}
\affiliation{${}^{y_m^2}$ Department of Mathematics, King’s College London, Strand, London WC2R 2LS, United Kingdom}
\affiliation{${}^{[y_m^2]_{2}}$ Deutsches Elektronen-Synchrotron DESY, Notkestraße 85, 22607 Hamburg, Germany}

\emailAdd{julius.julius@phys.ens.fr}
\emailAdd{nika.sokolova@desy.de}

\abstract{
In this paper, we discuss the construction of a map between weak (gauge) and strong (string) coupling degrees of freedom for the supersymmetric Wilson line-defect in the planar $\mathcal{N} = 4$ Super-Yang-Mills. 
By 
analysing the Partition Functions at zero and infinite coupling, 
we propose a map from degrees of freedom capturing single- and singlet two-particle states at zero coupling to 
infinite coupling.
This map predicts that the dimension of states in these particular sectors doubles as it goes from zero to infinite coupling. 
We test this prediction against the non-perturbative spectrum of insertions on the Wilson line obtained using integrability. 
In addition to already available integrability-based results,
we obtain the non-perturbative scaling dimensions of the simplest non-trivial operators with transverse spin about the Wilson line,
thereby extending the Quantum Spectral Curve construction to such charged sectors.
}

\begin{document} 
\begin{flushright}DESY-26-019\end{flushright}

\maketitle

\flushbottom

\input{intro_v2.tex}

\input{setup.tex}

\input{analysis.tex}

\input{map.tex}

\input{qsc_v2.tex}

\input{pade_v2.tex}

\input{disc_v2.tex}

\acknowledgments

We thank Costas Bachas, Abhijit Gadde, Barak Gabai, Victor Gorbenko, Nikolay Gromov, Christopher Herzog, Shiraz Minwalla, Sameer Murthy, Bendeguz Offertaler, Onkar Parrikar, Miguel Paulos, Elli Pomoni, Volker Schomerus and Gerard Watts for useful discussions. 
We thank Bendeguz Offertaler for carefully reading the manuscript and insightful comments thereon.

JJ would like to thank the Isaac Newton Institute for Mathematical Sciences, Cambridge, for support and hospitality during the programme The Many Faces of Boundaries, Impurities, and Defects, where some work on this paper was undertaken. 
This work was partially supported by EPSRC grant EP/Z000580/1.
JJ acknowledges the support of the
Marie Sk{\l}odowska-Curie Actions Staff Exchange Grant Number 
101182937 -- High-energy Intelligence
that enabled a visit to the
\'Ecole Polytechnique F\'ed\'erale de Lausanne (EFPL), where 
part of this work was conceptualised
during discussions with
Barak Gabai, Victor Gorbenko, Bendeguz Offertaler and Miguel Paulos.
JJ thanks the EPFL and Tata Institute for Fundamental Research for hospitality and for creating a stimulating research environment during his visits.
The work of JJ is supported by the European Union (ERC, FUNBOOTS, project number 101043588).
JJ expresses gratitude to the people of India for their continuing support towards the study of basic sciences. 

The work of NS is funded by the German Research Foundation DFG – SFB 1624 – “Higher structures, moduli spaces and integrability” –
506632645. Part of the work of NS was supported by
the European Research Council (ERC) under the European Union’s Horizon 2020 research and innovation program – 60 – (grant agreement No. 865075) EXACTC.

\appendix

\input{supermultiplets.tex}

\input{pade_adx.tex}

\bibliographystyle{JHEP.bst}
\bibliography{references}

\end{document}

%% file: intro_v2.tex
\section{Introduction}\label{sec:intro}

Dualities are ubiquitous in modern theoretical physics. One of the most famous of these is the AdS/CFT correspondence \cite{Maldacena:1997re, Gubser:1998bc,Witten:1998qj}, which predicts a holographic duality between string theory on the curved AdS background 
and a conformal field theory (CFT) defined on the boundary of the AdS space. 
Thus, the gauge-field degrees of freedom are mapped to the string degrees of freedom. Other examples of dualities include S- and T-dualities \cite{Montonen:1977sn,Sen:1994fa, Buscher:1987qj, Buscher:1987sk}, as well as identities such as the KLT-relations \cite{Kawai:1985xq}, which map gravitational and gauge theory amplitudes at tree level.

There is no reason to expect that repackaging of gauge degrees of freedom into string degrees of freedom would be smooth/perturbative, in the sense of being able to identify the alphabets parametrising the respective degrees of freedom using some simple map.
In principle, one can expect that the gauge theory degrees of freedom mix in some essentially non-perturbative way, and then re-emerge as string degrees of freedom.

Recently, the point of view of there existing a smooth map between gauge and string degrees of freedom was taken in~\cite{Gabai:2025hwf,FluxTubePade}.
Here the authors studied flux tubes in pure Yang-Mills theory placed on an AdS background. 
This setup offers dual descriptions in terms of gauge fields as well as an effective string theory.
The authors gave evidence for there being a smooth way to map the degrees of freedom between either end of the duality.

Inspired by these developments, in
this paper, we study a supersymmetric Wilson line-defect inserted in planar $\mathcal{N}=4$ Super-Yang-Mills (SYM) theory~\cite{Maldacena:1998im}. 
This example serves as an example of a defect, for which the degrees of freedom at weak (gauge) and strong (string) coupling are well understood~\cite{Giombi:2017cqn, Liendo:2018ukf, Ferrero:2021bsb, Ferrero:2023znz}, together with perturbative and analytical bootstrap approaches \cite{Barrat:2021tpn, Barrat:2022eim, Ferrero:2023gnu, Artico:2024wnt, Artico:2024wut} and for which the intermediate coupling regime is also accessible via integrability~\cite{Gromov:2012eu, Gromov:2013qga, Gromov:2015dfa, Grabner:2020nis,Julius:2021uka,Cavaglia:2021bnz,Cavaglia:2024dkk}. A combination of integrability and conformal bootstrap also leads to powerful numerical bounds for the operator product expansion (OPE) coefficients in this theory~\cite{Cavaglia:2021bnz, Cavaglia:2022qpg, Cavaglia:2022yvv, Cavaglia:2024dkk}. 
Thus, due to its solvability, it functions as a ``Rosetta Stone'' for Wilson line-defects, hopefully allowing to draw universal conclusions.

The degrees of freedom at weak and strong coupling give us two alphabets to construct states in either regime.
The central question that we try to understand in this paper is whether there is a smooth map between the alphabets at weak and strong coupling. 

By focusing on a certain sub-sectors of states in this theory which we call ``single-'' and singlet ``two-particle'' states respectively,
we propose a simple map between words in either alphabet. 
To test the map, we see that given a single-particle state, or a singlet two-particle state with some scaling dimension at zero coupling, the map predicts what the dimension of this state doubles at infinite coupling. 
We are able to verify this prediction using existing integrability-based results as well as by generating new ones. 
In particular, we extend the Quantum Spectral Curve (QSC) describing states in the supersymmetric Wilson line-defect~\cite{Grabner:2020nis,Julius:2021uka,Cavaglia:2021bnz,Cavaglia:2024dkk} to states with non-zero transverse spin about the Wilson line, to obtain the non-perturbative scaling dimension of some lowest lying bosonic spinning operators.

Thus, we are able to provide some more evidence towards the existence of a smooth map of degrees of freedom. 
In particular our message could be interpreted in the context of dualities between theories where there is no integrability, but which nevertheless allow for perturbative computations, as a
suggestion to proceed with a simple map between degrees of freedom, and use methods like Pad\'e approximation to generate non-perturbative predictions.

In this spirit, we also analyse the spectrum of two-particle states using Pad\'e approximation, even though the exact result is known. 
This serves two purposes. 
Firstly, we are able to benchmark the Pad\'e approximants against the exact result.
Secondly, since our map predicts the dimension of a two-particle state at strong coupling, this information can now potentially be combined with perturbative results at weak coupling into a double-sided Pad\'e approximation, which could more efficiently generate predictions for spectral data at intermediate coupling, that may be fed into flat-space $S$-matrix bootstrap approaches~\cite{Ghosh:2025sic}.

Our paper is organised as follows. In Section \ref{sec:setup}, we describe our main setup, introduce the alphabets capturing the degrees of freedom at weak (gauge) and strong (string) coupling, and review the procedure to count states in either regime. 
In Section~\ref{sec:refinedPF}, 
we analyse the Partition Functions at weak and strong coupling by refining it in various ways, and use this to draw predictions about the structure of the states in either regime. 
In Section~\ref{sec:mapalphabets}, we propose a map from degrees of freedom contributing to single- and singlet two-particle Partition Functions at weak coupling to the ones at strong coupling. 
This map predicts that the scaling dimensions of states double from zero to infinite coupling. 
In Section~\ref{sec:integrability}, we test this prediction by comparing with non-perturbative spectral data obtained from 
integrability for single- and two-particle states, and find a match.
Then, in Section~\ref{sec:pade}, we explore a possible application of our map by performing a Pad\'e analysis.
Finally, in Section~\ref{sec:discussion}, we discuss the conceptual and practical implications of such a map between weak (gauge) and strong (string) coupling degrees of freedom.

%% file: setup.tex
\section{Setup}
\label{sec:setup}
In this section we will provide the description of the supersymmetric Wilson line-defect in $\mathcal{N}=4$ SYM.
We will describe the setup at weak and strong coupling and introduce the degrees of freedom that capture the dynamics in either setting.
Finally we will review the procedure to enumerate states.

\subsection{Description}
\label{sec:desc}

Consider a supersymmetric (infinite-straight) Wilson line in $\mathcal{N} = 4$ SYM \cite{Maldacena:1998im}:
\begin{align}
    \mathcal{W} \coloneqq \mathrm{tr}\,\mathrm{P}\, \mathrm{exp} \,W_{-\infty}^{+\infty} 
    = \left[\mathrm{tr}\,\mathrm{P}\,\mathrm{exp} \int_{-\infty}^{+ \infty} dt \left[ i\, A_{t} + \Phi_{||}\right] \right]\;,
\end{align}
where $\text{P}$ denotes path-ordering along the line. 
In addition to the gluon, the supersymmetric Wilson line also includes a coupling to one of the six real scalars of the theory, which we denote as $\Phi_{||}$.

It is possible to make adjoint insertions along the Wilson line. In particular, consider the $n$-point insertion
\begin{align}
    \langle \langle \mathcal{O}_1(t_1)\dots \mathcal{O}_n(t_n)\rangle\rangle \coloneqq 
    \frac{\mathrm{tr}\left[\mathrm{P} 
    W_{-\infty}^{t_1}
    \mathcal{O}_1(t_1)  W_{t_1}^{t_2}
    \mathcal{O}_2(t_2)  \dots \mathcal{O}_n(t_n) W_{t_n}^{+\infty} \right]
    }{\langle \mathcal{W} \rangle}\;.
\end{align}
It can be shown~\cite{Drukker:2006xg} that this observable satisfies the axioms of an $n$-point function in a one-dimensional CFT and therefore can be treated as such.
In this paper, we are going to be interested in two-point functions in this one-dimensional defect CFT:
\begin{gather}
    \langle \langle \mathcal{O}(t_1) \mathcal{O}(t_2) \rangle \rangle = \frac{1}{(t_1 - t_2)^{2\,\Delta_{\mathcal{O}}}}\;,
\end{gather}
which are characterised by the scaling dimension $\Delta_{\mathcal{O}}$ of the insertion $\mathcal{O}$.

Due to insertion of the Wilson line-defect, the symmetries of the bulk theory $\mathcal{N}=4$ SYM are broken \cite{Liendo:2018ukf, Ferrero:2021bsb, Ferrero:2023znz}. 
The four-dimensional conformal group breaks to a one-dimensional conformal group along the Wilson line, and three-dimensional rotations about the Wilson line:
\begin{align}
    \mathrm{SO}(4,2)\to\mathrm{SO}(1,2)\times (\mathrm{SO(3)} \cong \mathrm{SU}(2))
    \;.
\end{align}
Due to the fact that we chose a particular real scalar $\Phi_{||}$ to couple to the Wilson line, the $R$-symmetry is also broken:
\begin{align}
    \mathrm{SO}(6) \cong \mathrm{SU}(4) \to \mathrm{SO}(5) \cong \mathrm{Sp(4)}
    \;.
\end{align}
The insertion of the Wilson line also breaks half of the supersymmetries and we have
\begin{align}
    \mathrm{PSU}(2,2|4)\to \mathrm{OSp}(4^*|4)\;.
\end{align}
As such, we have a one-dimensional defect superconformal field theory (1D dSCFT) that lives on the supersymmetric Wilson-line in $\mathcal{N}=4$ SYM.

States in this theory are generated by making insertions on the Wilson line and they are labelled by their quantum numbers under the bosonic part of the defect symmetry group:
\begin{align*}
    \begin{split}
        \Delta\;,\quad&\text{the scaling dimension labelling the 1D conformal group }\mathrm{SO}(1,2)\;,\\
        [j]\;,\quad&\text{the transverse spin labelling the } \mathrm{SO}(3) \text{ global rotations around the line}\;,\\
        (m,n)\;,\quad&\text{the Dynkin labels of the defect {\it R}-symmetry group }\mathrm{Sp}(4)\;.
    \end{split}
\end{align*}
Put together, we use the notation $[j]_\Delta^{(m,n)}$ to specify the labels that uniquely identify a state.
The labels $[j]$ and $(m,n)$ are non-negative integers, however the label $\Delta$ can be a non-trivial function of the parameters of the theory. 
We will be interested in the supersymmetric Wilson line in planar $\mathcal{N}=4$ SYM, and therefore the only free parameter we have is the 't Hooft coupling $\lambda\coloneqq g_{\texttt{YM}}^2\,N$.
Thus, $\Delta(\lambda)$ can in principle at least be a non-trivial function.

\paragraph{Supersymmetry and super-multiplets.}
A classification of super-conformal line-defects in various spacetime dimensions, was done in~\cite{Agmon:2020pde}.
This included a classification of all possible types of super-multiplets in these defect theories.
The explicit construction of the supermultiplets in the 1D dSCFT on the supersymmetric Wilson line in $\mathcal{N} = 4$ SYM was carried out in~\cite{Ferrero:2023znz}.
The results of both constructions are presented in Appendix~\ref{app:supermultiplets}, for the reader's convenience.
Importantly, there are three types of super-multiplets: short $\mathcal{B}$-type super-multiplets, semi-short $\mathcal{A}$-type super-multiplets, and long $\mathcal{L}$-type super-multiplets. 

There are two holographically dual descriptions of the 1D dSCFT, at weak and strong 't Hooft coupling.
The degrees of freedom in either description are captured by the super-multiplets.
We will describe how below.

\paragraph{Weak coupling.}
At weak coupling, the fields of $\mathcal{N}=4$ SYM rearrange themselves into super-multiplets, based on their quantum numbers under the symmetries preserved by the defect. In particular we have~\cite{Cooke:2017qgm}
\begin{align*}
    \begin{split}
        \Phi_{I}
        &\to \Phi_m \oplus \Phi_{||}\;,
        \\
        \Psi_{A\;\alpha} \pm \dot{\Psi}_{A\;\dot{\alpha}}
        &\to \Psi^{+}_{a\;\alpha} \oplus \Psi^{-}_{a\;\alpha}\;,
        \\
        \mathcal{F}_{\mu\nu}
        &\to \mathcal{F}_{ti}\oplus\mathcal{F}_{ij}\;,
        \\
        \mathcal{D}_\mu &\to \mathcal{D}_i \oplus \mathcal{D}_t\;.
    \end{split}
\end{align*}
Here, $I$ is an $\mathrm{SO}(6)$ vector index which breaks into $\mathrm{SO}(5)$ with vector index $m$; $A$ is an $\mathrm{SU}(4)$ fundamental index which breaks into an $\mathrm{Sp}(4)$ fundamental index $a$; $\alpha,\dot{\alpha}$ are an $\mathrm{SU}(2)$ fundamental indices; and $\mu,\nu$ are a $\mathrm{SO}(4)$ vector indices which break into  $\mathrm{SO}(3)$ vector indices $i,j$. 
The subscripts $||$ and $t$ are neutral under $\mathrm{SO}(5)$ and $\mathrm{SO}(3)$ respectively.
Only the derivative $\mathcal{D}_t$ generates conformal descendents. 

At weak coupling, it was shown~\cite{Ferrero:2023znz} that there are infinitely many multiplets that contribute, \textit{cf.}~equation~\eqref{eqn:zMultContent}.
The simplest of these is a $\mathcal{B}_1$-type multiplet with super-primary quantum number $[0]_1^{(0,0)}$. 
The super-primary can be identified $\Phi_m$, and the super-descendents can be mapped to one of the fermions above and the combination $\mathcal{F}_{it} + \mathcal{D}_i\Phi_{||}$~\cite{Cooke:2017qgm}.
Another simple multiplet is of $\mathcal{A}_2$-type multiplet with super-primary quantum number $[0]_1^{(0,0)}$. The super-primary can be identified as $\Phi_{||}$, with the other fermion and $\mathcal{F}_{ij}$ being identified as some of the super-descendents~\cite{Cooke:2017qgm}.
The rest of the infinitely many super-multiplets can be thought of as
arising due to preparing states with by acting with the orthogonal derivative $\mathcal{D}_i$.

\paragraph{Strong coupling.}
At strong coupling, by the AdS/CFT correspondence, 
the setup is dual to an open string in the bulk, ending on the Wilson line in the boundary.
The open string world-sheet carves out an AdS${}_{2}$ inside the AdS${}_5$, and insertions on the Wilson line correspond to fluctuations on this AdS${}_{2}$ world-sheet~\cite{Giombi:2017cqn}.
There are three types of fluctuations:
\begin{itemize}
    \item 5 massless bosonic modes corresponding to $S^5$ fluctuations, denoted as $y_m$\;,\;$m = 1,\dots,5$\;,
    \item 3 massive bosonic modes corresponding to fluctuations on the AdS$_{3}$ orthogonal to the world-sheet, denoted as $f_i$\;,\;$i=1,\dots,3$\;,
    \item 8 fermionic modes, denoted as $\psi_{a\;\alpha}$\;,\;$a=1,\dots,4$\;,\;$\alpha=1,2$\;.
\end{itemize}
These fluctuations can be captured by a $\mathcal{B}_1$-type multiplet with super-primary quantum numbers $[0]_1^{(0,1)}$ identified with $y_m$, and the other types of fluctuations identified with the super-descendents. Conformal descendents are created by derivatives $\partial_t$ along the line.

This completes the description of the main setup.
In the next section, we will review how one can enumerate states in the 1D dSCFT, which can be built out of the alphabets that capture the degrees of freedom at weak and strong coupling.

\subsection{Review of State Counting}
\label{sec:counting}

We start with a general review of state counting. There are many resources available in the literature regarding this, we follow the presentation of~\cite{Gadde:2020yah}.
Then, we review how specifically this counting apparatus can be applied to enumerate states in the 1D dSCFT at weak and strong coupling. 
This computation was originally carried out in~\cite{Ferrero:2023znz}. 
Our review mostly follows the same method, though few technical details vary.

\paragraph{Building a partition function.} 
Let us start with the example of a theory where we have a single bosonic degree of freedom. 
We capture this using a single bosonic letter $\varphi$, and define the single-letter Partition Function (SLPF) to be
\begin{align}
    z_\texttt{B}(\varphi) \coloneqq \varphi\;.
\end{align}
The multi-letter Partition Function (MLPF), which captures all possible words that can be generated given a set of single-letters, is obtained by \emph{Plethystic  Exponentiation} (PE) of the SLPF. For bosonic single-letters, it is defined as
\begin{equation}
    \mathcal{Z}_\texttt{B}(\varphi) \coloneqq \mathrm{PE} [z_\texttt{B}(\varphi)] = e^{\sum_{n=1}^{\infty} \frac{1}{n} z_\texttt{B}(\varphi^n)} = \frac{1}{1 - \varphi} = 1 + \varphi + \varphi^2 + \varphi^3 + \dots\;.
\end{equation}
For the case of a single bosonic degree of freedom, it is clear how the PE generates the MLPF. 
Here we can see that we have a unique way to construct words with different number of letters, as expected in the case of one bosonic letter. 
This procedure is easily generalised if we have multiple bosonic letters. For example, if we have two letters $\varphi$ and $\varrho$, then
\begin{align}
    z_\texttt{B}(\varphi,\varrho) \coloneqq \varphi + \varrho\;,\quad \mathcal{Z}(\varphi,\varrho) = \frac{1}{1-\varphi}\frac{1}{1-\varrho}\;.
\end{align}
Next, consider the case where have a single fermionic degree of freedom, captured by the single-letter $\varsigma$ to give the SLPF
\begin{align}
    z_\mathtt{F} \coloneqq \varsigma\;.
\end{align}
Then one need to modify the PE to generate the MLPF. It is done in the following way:
\begin{equation}
    \mathcal{Z}_{\texttt{F}}(\varsigma) \coloneqq \widetilde{\mathrm{{PE}}} [z_\texttt{F}] = e^{-\sum_{n=1}^{\infty} \frac{1}{n} z_{\texttt{F}}((-1)^n\varsigma^n)} = 1 + \varsigma\;,
\end{equation}
which accounts the fact that we cannot have more that one fermionic letter in a word.

\paragraph{Incorporating global symmetries.}
In many physically interesting cases, 
the fields 
transform under irreps of some global symmetries.
To keep track of these symmetries, it is useful to dress each single-letter in the SLPF with the \emph{character} 
of the irrep of the global symmetry, 
under which the field corresponding to that single-letter transforms. 
For the example of a single bosonic degree of freedom, the SLPF is therefore
\begin{align}
    z_{\mathtt{B}}(\varphi;a_i) = \varphi\,\chi_{R}(a_i)\;,
\end{align}
where $\chi_{R}(a_i)$ is a character of a given irrep $R$ of the symmetry group $G$, $a_i$ are character variables/fugacities.
The MLPF is generated as usual by PE, we have
\begin{align}
    \mathcal{Z}_\texttt{B}(\varphi;a_i) = \mathrm{PE}[z_\texttt{B}(\varphi;a_i)] = 
    e^{\sum_{n=1}^{\infty} \frac{1}{n} z_\texttt{B}(\varphi^n;a_i^n)}
    \;.
\end{align}
To count multi-letter words in a specific representation, 
we need to project the tensor product onto that representation. 
This is done using the orthogonality of characters:
\begin{equation}
    \frac{1}{|W(G)|} \oint \prod_{i = 1}
^{r} \frac{d a_i}{2\pi i\, a_i} {\Delta}(a_i) \chi_{R}(a_i) \chi_{R'}(a_i) = \delta_{R R'}\;,
\label{orth}
\end{equation}
where $r$ is the rank of $G$, $|W(G)|$ is the cardinality of the associated Weyl group and ${\Delta}(a_i)$ is the Vandermonde determinant.
To illustrate, the projection of the MLPF of a single bosonic degree of freedom onto the global symmetry sector furnishing the irrep $R^\prime$ is
\begin{equation}\label{eqn:ProjectGlobalSYmmetries}
    \mathcal{Z}(\varphi)\bigg|_{R^\prime} = \frac{1}{|W(G)|} \oint \prod_{i = 1}^{r} \frac{d a_i}{2\pi i\,  a_i} \Delta(a_i) \mathcal{Z}(\varphi; a_i) \chi_{R^\prime}(a_i)\;.
\end{equation}
The expressions above are for continuous global symmetries.
A similar treatment can be made in order to capture discrete symmetries as well, as described in~\cite{FluxTubePade}\footnote{We thank Bendeguz Offertaler for sharing the projection formula with us.}.

\paragraph{Extracting the (super-)primary contribution.}
We are interested in counting super-primary states in the 1D dSCFT. As such, we need to understand how to project out conformal descendents and super-descendents from the MLPF.

In general, the counting procedure will output the MLPF $\mathcal{Z}_{\texttt{full}}(q;\dots)$, which contains all states:
super-primaries, super-descendent conformal primaries, and their conformal descendents.
It is expressed in terms of a fugacity $q$, which captures the scaling dimension of a multi-letter state, as well as other fugacities capturing global symmetries, denoted by the elepesis.

In order to extract the MLPF of conformal primary states only, 
recall that conformal descendents are generated from conformal primaries by the action of momentum generators, and this increases their dimension by unity. Specialising to 1D, which is the case that we are interested in, we have for the SLPF
\begin{equation}
  |z_\texttt{descendant} \rangle 
  =  P  | z_\texttt{primary} \rangle 
  = q |z_\texttt{primary} \rangle\;,
\end{equation}
where $P$ is the momentum generator.
Thus, we can construct all possible conformal descendants given a MLPF of solely conformal primaries $\mathcal{Z}_{\texttt{primary}}$ as follows:
\begin{equation}
    |\mathcal{Z}_\texttt{descendant} \rangle = (q + q^2 + q^3 + \dots) | \mathcal{Z}_\texttt{primary} \rangle = \frac{q}{1-q} | \mathcal{Z}_\texttt{primary} \rangle\;.
\end{equation}
The MLPF including both conformal primaries and descendents, $\mathcal{Z}_\texttt{full}$, can therefore be written as
\begin{equation}
 |\mathcal{Z}_\texttt{full} \rangle =    | \mathcal{Z}_\texttt{primary} \rangle + (q + q^2 + q^3 + \dots) |\mathcal{Z}_\text{primary} \rangle  = \frac{1}{1 - q} | \mathcal{Z}_\texttt{primary} \rangle\;.
\end{equation} 
Thus, to extract the MLPF of primary states, we just need to multiply by $(1 - q)$:
\begin{equation}\label{eqn:ExtractConfPrimaries}
    | \mathcal{Z}_\texttt{primary} \rangle = (1 - q) |\mathcal{Z}_\texttt{full} \rangle\;.
\end{equation}
There is an important subtlety that one should keep in mind when carrying out the above, which is the possibility of appearance of conformal null vectors.
Constraints arising from conformal null vectors need to be suitably accounted for in the definition of the SLPF itself.

A similar trick can be used to extract the MLPF of only super-conformal primaries. 
Super-descendent conformal primaries are obtained from super-conformal primaries by acting with supercharge operators $\mathcal{Q}_A$. 
Here $A$ is a label which accounts for the possibility of have multiple supercharges. 
Because supercharge operators are fermionic, we can act only once with them. Therefore, we have
\begin{equation}
 |\mathcal{Z}_{\text{primary}} \rangle = |\mathcal{Z}_\text{superprimary} \rangle + \prod_{A = 1}^{{\mathcal{N}}} (1 + \mathcal{Q}_A) | \mathcal{Z}_{\text{superprimary}} \rangle\;,
\end{equation}
where ${{\mathcal N}}$ is the number of supercharges\footnote{In our case, we act with eight supercharges listed in Table \ref{tab:supercharges}.}.
This suggests that the super-descendants can be factored out in the following way:
\begin{equation}\label{eqn:projectSuperDesc}
    |\mathcal{Z}_\texttt{super-primary} \rangle = \frac{1}{1 + \prod_{A = 1}^{{\mathcal N}} (1 + \mathcal{Q}_A)} | \mathcal{Z}_\text{primary} \rangle.
\end{equation}
One must keep in mind however, that there is a subtlety to take into account when factoring out super-descendents in this way. Notice that there is an implicit assumption in the passage above that all supercharges act on the super-primary to create viable states. In other words, strictly speaking, this factorisation procedure should work only for long super-multiplets. Therefore, we run the risk of mistakes in calculating the contribution semi-short and short multiplets. To handle this, recall the conditions under which a long super-multiplet splits into a short and a semi-short multiplet. The splitting happens when the scaling dimension of the super-primary of a long multiplet saturates the unitarity bound, leading to the appearance of a null state somewhere in the super-multiplet, and causing the split. The unitarity bound is usually saturated by the super-primary at the free-theory point. Once even a tiny amount of interactions is switched on, the scaling dimension of the super-primary goes strictly above the unitarity bound, causing recombination. Thus, if we keep in mind that we are doing our counting procedure just outside the free-theory point, then all semi-short and short multiplets that are generated by multiplet splitting can be considered as long multiplets. 

The only multiplets which remain to be considered are the ``absolutely protected'' ones. These are not generated by the super-primary of a long multiplet saturating the unitarity bound saturating the shortening conditions. They need to be removed by hand. Therefore, we have
\begin{gather}\label{eqn:ExtractSuperConfPrimaries}
    \mathcal{Z}_{\texttt{long super-primary}}
    =  \frac{1}{1 + \prod_{A = 1}^{{\mathcal{N}}} (1 + \mathcal{Q}_A)} \big[\mathcal{Z}_{\texttt{primary}}
    - \mathcal{Z}_{\texttt{absolutely protected}}
    \big]\;.
\end{gather}
It turns out that in the case of the 1D dSCFT, only one type of absolutely protected supermultiplets are present, and furthermore, each such multiplet has unit multiplicity~\cite{Liendo:2018ukf,Ferrero:2023znz}. Therefore, the subtraction is straightforward in this case.

\paragraph{Infinite coupling.}
\label{sec:strong}
In the strong coupling limit, as reviewed in Section~\ref{sec:desc},
the single-letters are the various fluctuations of a 
an AdS${_2}$ world-sheet~\cite{Giombi:2017cqn},
captured by a $\mathcal{B}_1$-type multiplet whose super-primary $y_m$ has quantum numbers $[0]^{(0,0)}_1$.
The content of this multiplet is
\begin{equation}
    y_m: [0]_{1}^{(0, 1)} 
    \to 
    \psi_{a\,\alpha} : [1]_{\frac{3}{2}}^{(1, 0)} 
    \to 
    f_i: [2]_{2}^{(0, 0)}\;.
\end{equation}
Lastly, we have the action of the derivatives $\partial_t$, which are uncharged under the continuous global symmetries, and
increase the dimension of a state by unity.

Therefore, the SLPF for the 1D dSCFT at infinite coupling is
\begin{align}
    \label{eqn:SLPFStrong}
    \begin{split}
        z_\infty(q;a,b;\ell) &= \frac{1}{1-q}\bigg[q\,\chi_{\mathrm{SP}(4)}^{(0, 1)}(a,b)\,\chi_{\mathrm{SO}(3)}^{[0]}(\ell)
        +
        q^{3/2}\,\chi_{\mathrm{SP}(4)}^{(1, 0)}(a,b)\,\chi_{\mathrm{SO}(3)}^{[1]}(\ell)
        +\\ 
        &+
        q^2\,\chi_{\mathrm{SP}(4)}^{(0, 0)}(a,b)(a,b)\,\chi_{\mathrm{SO}(3)}^{[2]}(\ell)
        \bigg]\;,\\
        &= \frac{1}{1-q}\bigg[
        q\,\left( \frac{a^2}{b} + \frac{b}{a^2} + b + \frac{1}{b} + 1 \right)
        + 
         q^{3/2}\,\left(\frac{a}{b} + \frac{b}{a} + a + \frac{1}{a} \right) \left( \ell + \frac{1}{\ell} \right)
         \\
         &+
         q^2\,\left( \ell^2 + \frac{1}{\ell^2} + 1\right)
        \bigg]\;.
    \end{split}
\end{align}
Here $q$ is again the fugacity associated with the dimension of a state, and 
the factor of $1/({1 - q})$ accounts all possible derivatives $\partial_t$ applied on a single-letter. 

To generate the MLPF, we must
\begin{enumerate}
    \item Apply PE to the SLPF~\eqref{eqn:SLPFStrong}, remembering to appropriately modify for the case of fermionic letters.
    \item Extract the MLPF of conformal primaries using equation~\eqref{eqn:ExtractConfPrimaries}.
    \item Compute the partition function of the absolutely protected supermultiplets.
    \item Extract the Partition Function of superconformal primaries using equation~\eqref{eqn:ExtractSuperConfPrimaries}.
    \item Project onto the desired continuous global symmetry representation using equation~\eqref{eqn:ProjectGlobalSYmmetries}.
\end{enumerate}
In our case, the only absolutely protected supermultiplets are those of $\mathcal{B}_1$-type with the super-primary quantum numbers $[0]_{m+n}^{(m,n)}$, subject to  $a\leq1$~\cite{Agmon:2020pde}. 
However, $\mathcal{B}_1$-type multiplets with super-primary quantum numbers $[0]_{1+n}^{(1,n)}$ do not contribute either at weak or at strong coupling~\cite{Ferrero:2023znz}. 
Furthermore, the $\mathcal{B}_1$-type multiplets with super-primary quantum numbers $[0]^{(0,m)}_m$ multiplets have unit multiplicity~\cite{Liendo:2018ukf,Ferrero:2023znz}. Therefore, we have
\begin{gather}
    \mathcal{Z}_\texttt{absolutelty protected} = \sum_{m=1}^\infty \chi_{[0]_m^{(0,m)}}
    = \sum_{m=1}^\infty q^m\,\chi^{(0,m)}_{\mathrm{SP}(4)}\,\chi_{\mathrm{SO(3)}}^{[0]}
    \;. 
\end{gather}
Plugging everything in and following the steps above, one gets~\cite{Ferrero:2023znz}
\begin{align}
\nonumber
    &\mathcal{Z}_{\infty}(q) = q^2 + 2\,q^4 + 4\, q^6 + q^7 + 9\,q^8 + 5\,q^9 + 21 \,q^{10} + 20\, q^{11}+ 55\, q^{12} + 65\, q^{13} + \\
    & + 149\, q^{14} + {207\, q^{15}+ 416\, q^{16}} +\mathcal{O}(q^{17}).
    \label{countingstong}
\end{align}
for the singlet sector, \textit{i.e.} the MLPF of states which are singlets under the continuous global symmetries $\mathrm{SP}(4)\times\mathrm{SO}(3)$.
The exponent of $q$ gives the scaling dimension $\Delta_\infty$ at zero coupling.

\paragraph{Zero coupling.}
\label{sec:weak}

At weak coupling, the SLPF for the 1D SCFT is given by
\begin{align}\label{eqn:SingleLetterWeak}
    z_0 = w + \frac{1}{1-q}z_{\texttt{multiplets}}\;.
\end{align}
Here, we have introduced the letter $w$ associated with the Wilson line, the reason for which is explained below; and $z_{\texttt{multiplets}}$ captures the characters of all the super-multiplets that contribute to the SLPF. 

In the weak coupling picture, we are considering the insertion of colour-charged fields on a Wilson line, and as such this
colour structure of the fields needs to be taken into account. 
In particular, the fields transform under the adjoint of the gauge group $\mathrm{SU}(N)$, and we only want to enumerate multi-letter states which are gauge invariant. 
Furthermore, we are interested in the large-$N$ limit and therefore we cannot use the the orthogonality relation (\ref{orth}), as it would introduce an infinite number of character variables.
There is a known procedure how to count multi-trace and single-trace operators developed for $\mathcal{N}=4$ SYM \cite{Sundborg:1999ue, Aharony:2003sx, Kinney:2005ej, Benvenuti:2006qr, Feng:2007ur, Forcella:2007wk, Mandal:2006tk}, which can be adapted for the case of the Wilson line as we describe below.

\begin{figure}[htbp]
  \centering
  \begin{tikzpicture}[scale=1.0]
    \def\R{1.4} 
    \draw[very thick, green!65!black,
          postaction={decorate},
          decoration={
            markings,
            mark=at position 0.25 with {\arrow{Stealth[length=3.5mm]}},
          }
    ] (0,0) circle (\R);
    \coordinate (Op) at (90:\R);
    \fill[violet!75!black] (Op) circle (2.8pt);
    \node[above=4pt, violet!75!black, font=\small] at (Op) {$O_{\text{inserted}}$};
    \node[green!65!black, font=\small] at (0:\R+0.25) {$w$};
  \end{tikzpicture}
    \caption{Inserted operators $O_\text{inserted}$ in the Wilson line can be seen as a single-trace operator where one of single letters is a Wilson line (only one Wilson line letter is allowed).}
    \label{CountingWilson}
\end{figure}
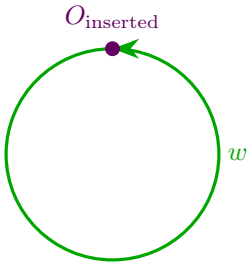

To enumerate single-trace states 
in an $\mathrm{SU}(\infty)$ gauge theory, given the SLPF $z(q)$, one needs to modify the PE as
~\cite{Bianchi:2006ti, Dolan:2007rq}:
\begin{equation}
    \mathcal{Z}_{\mathrm{SU}(N)\text{ }\texttt{single-trace}}(q) = \sum_{n = 2}^{\infty} \frac{1}{n} \sum_{d | n} \phi(d) z(q^d)^{n/d} = - \sum_{d = 1}^{\infty} \frac{\phi(d)}{d}\mathrm{log} (1 - z(q^d)) - z(q)\;,
\end{equation}
where $\phi(d)$ is the Euler totient function.
In our case, we have 
the
configuration of operators inserted into the Wilson line. 
However, if we recall that the Wilson line can be conformally mapped to a circle, 
then it is easy to convince oneself that we just need to count single-trace operators where one operator is the Wilson line itself, \textit{cf.}~Figure \ref{CountingWilson}. 
This is the reason behind introducing a new letter $w$ in the single-letter partition function~\eqref{eqn:SingleLetterWeak} representing a Wilson line. 
The required MLPF is 
\begin{align}
    \mathcal{Z}_{\texttt{insertions}}(q; a, b; \ell) = \partial_{w}\  \mathcal{Z}_{\mathrm{SU}(N)\text{ }\texttt{single-trace}}(w;q; a, b; \ell)\bigg|_{w=0}\;, 
\end{align}
which now counts all multi-letter insertions at a point on the Wilson line.

It remains to give the form of $z_\texttt{multiplets}$.
By expanding the $\mathcal{N}=4$ SYM SLPF suitably and suitably restricting the character fugacities to capture the continuous global symmetries preserved by the Wilson line-defect, the authors of~\cite{Ferrero:2023znz}
extracted the contribution of various super-multiplets to the SLPF in the weak coupling limit.
Their result is~\cite{Ferrero:2023znz}
\begin{align}\label{eqn:zMultContent}
   z_{\texttt{multplets}} = 
   \chi_{[0]^{(0,1)}_1} +
   \sum_{j=0}^\infty \chi_{[2j]^{(0,0)}_{j+1}}\;.
\end{align}
Here $[0]_1^{(0,1)}$ gives the quantum numbers of the super-primary $\Phi_m$ of a $\mathcal{B}_1$-type multiplet, $[0]_1^{(0,0)}$ gives the quantum numbers of the super-primary $\Phi_{||}$ of an $\mathcal{A}_2$-type multiplet, and the rest are $\mathcal{A}_1$-type multiplets.  
The SLFP at weak coupling is different from the strong coupling case.

We can now proceed analogously to the strong coupling case. 
Doing so yields the result~\cite{Ferrero:2023znz}
\begin{align}
\nonumber
  & \mathcal{Z}_{0}(q) = q + 2\, q^2 + 6\, q^3 + 25\, q^4 + 128\, q^5 + 758\, q^6 + 4986\, q^7 + 35\, 550\, q^8 + \\
  \nonumber
  & + 270\, 289\, q^9 + 2\, 166\, 106\, q^{10} + 18\, 140\,558\, q^{11} + 157\, 713\, 769\ q^{12} + 1\, 416\, 049\, 592\,q^{13} +  \\ & \nonumber   +13\, 075\, 731\, 238\, q^{14} + 123\,752\,918\,970\, q^{15} + 1\,197\,087\,829\,466 q^{16} + 11\,807\,452\,628\,689\, q^{17} \\ & + 118\,517\,478\,250\,082\, q^{18} + 1\,208\,558\,309\,932\,966\, q^{19}
  + 12\,501\,949\,485\,483\,041\,q^{20}
  +\mathcal{O}(q^{21})\;, 
  \label{countingweak}
\end{align}
for the singlet sector, \textit{i.e.} the MLPF of states which are singlets under the continuous global symmetries $\mathrm{SP}(4)\times\mathrm{SO}(3)$.
The exponent of $q$ gives the scaling dimension $\Delta_0$ at zero coupling.

%% file: analysis.tex
\section{Analysis of Partition Functions}
\label{sec:refinedPF}

In this section, we will analyse the MLPFs of the 1D dSCFT at zero and infinite coupling, and draw some conclusions as to the composition of multi-letter states.

\subsection{Refining the State Counting}
\label{sec:refined}

It is possible to further refine the MLPF to study various properties of states that we may be interested in.

Generically a multi-letter state gets contributions elements of various super-multiplets, as well as derivatives.
We define a single-particle state as that which is composed of a single element of one of the supermultiplets that form the SLPF.  
Higher-particle words can be defined in a similar way: an $n$-particle word is defined as a word composed of $n$ elements from the supermultiplets that form the SLPF.
Of course, this definition doesn't count the derivatives $\mathcal{D}_t$ or $\partial_t$, and therefore, an $n$-particle word may contain any number thereof. 
It is easy to keep track of the particle-number of words in the MLPF, by including an extra fugacity to capture whether the contribution to the word comes from an element of a super-multiplet or from a derivative. 
This is done by modifying the SLPF $z(q)$ as~\cite{Ferrero:2023znz}
\begin{align}
    z(q;l) \coloneqq l\,z(q)\;.
\end{align}
After going through the plethystic procedure, the MLPF will now be expressed in terms of both the fugacity $q$, which captures the scaling dimension, and additionally, the fugacity $l$, which captures the particle number.
To illustrate, under this refinement, the singlet MLPF at 
weak and 
strong coupling becomes~\cite{Ferrero:2023znz}
\begin{align}
    \label{eqn:ZWeakSingletParticles}
    \mathcal{Z}_0(q;l) &=
    l q+ 2 l^2 q^2+\left(4 l^3+2 l^2\right) q^3+\left(10 l^4+12 l^3+3 l^2\right) q^4
    +\dots
    \;,\\
    \label{eqn:ZStrongSingletParticles}
    \mathcal{Z}_{\infty}(q;l)
    &=
    l^2 q^2+\left(l^4+l^2\right) q^4
    +\left(l^6+2 l^4+l^2\right) q^6 
    + l^4 q^7
    + \left(l^8+3 l^6+4 l^4+l^2\right) q^8 + \dots\;.
\end{align}
This refinement already gives us some information. 
In particular, by comparing the discrepancy dimension of a word and the number of particles, we can draw conclusions on whether a multi-letter word got contributions from super-descendent or derivative single-letters.
We will analyse this below.

\subsection{Contributions of Derivatives and Super-descendents}
\label{sec:refineddQdt}
In order to keep track of the fine structure of a super-primary state to see if,
apart from having contributions from super-primary single-letters of various multiplets, it also
gets contributions from single-letters due to derivatives and super-descendents,
we introduce extra fugacities $dt$ and $dQ$.

\paragraph{Keeping track of derivatives.}
The contribution of derivatives to the SLPF is given in the factor $\frac{1}{1-q}$ that pre-multiplies it. 
One introduces the $dt$ fugacity here, 
\begin{align}
    z(q) \equiv \frac{1}{1-q}\bigg[\dots\bigg] \to  
    z(q;dt) \coloneqq \frac{1}{1 - dt\,q} \bigg[\dots\bigg]
    \;.
\end{align}
Then, one follows the plethystic procedure as usual, but remember to also modify the formula~\eqref{eqn:ExtractConfPrimaries} to
\begin{equation}\label{eqn:ExtractConfPrimariesdt}
    | \mathcal{Z}_\texttt{primary} \rangle = (1 - dt\,q) |\mathcal{Z}_\texttt{full} \rangle\;.
\end{equation}
while extracting the conformal primary contribution. To illustrate, here is the singlet MLPF at infinite coupling with the $dt$ fugacity switched on
\begin{align}
    \mathcal{Z}_{\infty}(q;l;dt) = 
    l^2 q^2
    +q^4 \left({dt}^2 l^2+l^4\right)
    +q^6 \left({dt}^4 l^2+2 {dt}^2 l^4+l^6\right)
    +\dots
    \;.
\end{align}

\paragraph{Keeping track of super-descendents.}
We use a similar trick to keep track of super-descendent contribution to a multi-letter word. 
However, in this case, the SLPF of each supermultiplet needs to be modified. 
It is done in the following way.
The term corresponding to the super-primary is kept unchanged.
The term corresponding to a super-descendent at level $k$, gets multiplied by $dQ^k$.
Finally, the projection formula~\eqref{eqn:projectSuperDesc} gets modified to 
\begin{equation}\label{eqn:projectSuperDescdQ}
    |\mathcal{Z}_\texttt{super-primary} \rangle = \frac{1}{1 + \prod_{A = 1}^{{\mathcal N}} (1 + dQ\,\mathcal{Q}_A)} | \mathcal{Z}_\text{primary} \rangle.
\end{equation}
Carrying out this procedure, we see that the first non-trivial contribution from a super-descendent to the singlet MLPF at infinite coupling occurs at $\mathrm{Q}(q^8)$. We get
\begin{align}\label{eqn:strongSingletdQOq8}
    \mathcal{Z}_{\infty}(q;l;dt;dQ)\bigg|_{q^8}  
    =
    \left({dt}^6 l^2 +4 {dt}^4 l^4 +2 {dt}^2 l^6 +{\color{red} dQ^4} l^6 +l^8\right)q^8
    \;.
\end{align}
We extracted it also at $\mathrm{Q}(q^{10})$, we get
\begin{align}\label{eqn:strongSingletdQOq10}
    \begin{split}
    \mathcal{Z}_{\infty}(q;l;dt;dQ)\bigg|_{q^{10}}  
    &=
    \left({dt}^8 l^2
    +6 {dt}^6 l^4
    +{\color{red} dQ^2} {dt}^4 l^5
    +5 {dt}^4 l^6
    +2 {\color{red} dQ^2} {dt}^3 l^6
    +2 {\color{red} dQ^4} {dt}^2 l^6
    \right.
    \\&
    \left.
    +2 {dt}^2 l^8
    +{\color{red} dQ^4} l^8
    +l^{10}\right)q^{10}
    \;.
    \end{split}
\end{align}
The terms with $dQ$ fugacities raised to some power, which have also been highlighted in red in the above two expressions should be interpreted in a particular way. To illustrate, consider the term ${\color{red} dQ^4} l^6q^8$, it must be interpreted as
\begin{equation}
    \label{eqn:RuleSingleParticle}
    \begin{minipage}{\textwidth}
    \begin{displayquote} 
    {\it
    ${\color{red} dQ^4} l^6q^8$ : 
    A word with dimension $\Delta_\infty = 8$, created by 6 single-letters taken from the $\mathcal{B}_1$-type multiplet with super-primary quantum numbers $[0]_1^{(0,0)}$. 
    We act on the set of six super-primary single-letters in all possible ways with 4 supercharges, to generate the possible sets of six single-letters that can be used to build the word.
    A single linear combination of these words contributes to the super-primary state.
    }
    \end{displayquote}
    \end{minipage}
\end{equation}

We can use this refinement to extract some information about  
super-primaries that are charged  under $\mathrm{SO}(3)$
at infinite coupling. 
We focus on those which are singlets of $\mathrm{Sp}(4)$, and transform in the irrep $[2j]$ of $\mathrm{SO}(3)$, with $j\geq1$.
We isolate only the lowest lying state for each value of $j$. We find
\begin{align}
    \begin{split}
        \label{eqn:SO3ChargedGroundState}
        \mathcal{Z}_{\infty;[2]}(q;dt;dQ)
        &=
        {dQ}^2 l^3 q^4+\dots\;, \\
        \mathcal{Z}_{\infty;[4]}(q;dt;dQ)
        &=
        {dQ}^4 l^4 q^6+\dots\;, \\
        \mathcal{Z}_{\infty;[6]}(q;dt;dQ)
        &=
        {dQ}^6 l^5 q^8+\dots\;. 
    \end{split}
\end{align}
We conclude firstly that the lowest lying state is unique, and it has dimension $\Delta_\infty = 2(j+1)$.
Secondly, 
we see that this state is composed of $(j+2)$-particles with the action of $2j$ supercharges.
We have checked this explicitly until $\Delta_\infty=16$.

\subsection{Two-particle Words}
Let us use the procedure of refining the MLPF by particle-number, to 
restrict to only two-particle words at weak and strong coupling, we get~\cite{Ferrero:2023znz}
\begin{align}
    \mathcal{Z}_{0}(q;l)\bigg|_{l^2} &= \sum_{n=2}^\infty \bigg(1+\left\lfloor\frac{n}{2}\right\rfloor\bigg)l^2q^n = 
    l^2\left(2q^2 + 2q^3 + 3q^4 + 3q^5 + 4q^6 + 4q^7 +\dots\right) \;,\\
    \mathcal{Z}_{\infty}(q;l)\bigg|_{l^2} &= l^2\sum_{n=1}^\infty q^{2n} = l^2\left( q^2 + q^4 + q^6 + \dots \right)
    \;.
\end{align}
We see that the growth of two-particle words is much slower that the total growth of states at weak~\eqref{countingweak} and strong~\eqref{countingstong} coupling. 
This is especially drastic at weak coupling, with one new state added at every even integer value of $\Delta_0$.

We further refine the SLPF to keep track of contributions from specific super-multiplets:
\begin{align}
    z_{\texttt{multiplets}}(q;l_j) =
    l_1\,\chi_{[0]^{(0,1)}_1} + \sum_{j=0}^\infty l_{2+j}\, \chi_{[2j]^{(0 ,0)}_{j+1}} 
    \;.
\end{align}
The MLPF for two-particle singlet words is
\begin{align}\label{eqn:2particleMLPFrefined}
    \mathcal{Z}_{0}(q;l_j)\bigg|_{l_j^2} = 
    \sum_{n=2}^\infty q^n\sum_{i=0}^{\lfloor\frac{n}{2}\rfloor}l_{i+1}^2
    \;,
\end{align}
which we have explicitly checked by expanding the PE up to $\Delta_0 = 12$.
Expanded, it gives
$$
    Z_{0}(q;l_j)\bigg|_{l_j^2} = 
    (l_1^2 + l_2^2)q^2 + (l_1^2 + l_2^2)q^3 
    + (l_1^2 + l_2^2 + l_3^2)q^4 + (l_1^2 + l_2^2 + l_3^2)q^5 + \dots
    \;.
$$
We immediately conclude
that all the two-particle singlet words are built from taking two elements of the same super-multiplet.

Below we will analyse it and other MLPFs 
further by switching on fugacities with respect to $dt$ and $dQ$.

\subsection{Towers and Trees of Derivatives}
\label{sec:TowerTree}
We begin by studying the two-particle singlet MLPF at zero coupling.
Then we analyse various MLPFs at infinite coupling.

\paragraph{Two-particle states at zero coupling.}
Switching on the $dt$ and $dQ$ fugacities in~\eqref{eqn:2particleMLPFrefined}, we get
\begin{align}\label{eqn:2particleMLPFrefineddt}
    \mathcal{Z}_{0}(q;l_j;dt;dQ)\bigg|_{l_j^2} = 
    (l_1^2 + l_2^2)q^2 + (dt\,l_1^2 + dt\,l_2^2)q^3 
    + (dt^2l_1^2 + dt^2l_2^2 + l_3^2)q^4 
    + \dots
    \;.
\end{align}
Let us unpack the information from this formula.
We start with 2 two-particle words at $\Delta_0 = 2$ generated by the super-multiplets associated with the fugacities $l_1$ and $l_2$ respectively.
From then onwards,
a new super-multiplet contributes a two-particle word at every even integer.
Once a two-particle word has been created, it functions as a seed to germinate an entire tower of two-particle words, obtained by acting on it with the derivative $\mathcal{D}_t$.

\paragraph{Multi-particle states at infinite coupling.}
Similarly, we can analyse the two-particle singlet MLPF at infinite coupling, with the $dt$ and $dQ$ fugacities switched on. We get
\begin{align}\label{eqn:2particleStrongMLPFdtdQ}
    \mathcal{Z}_{\infty}(q;l;dt;dQ)\bigg|_{l^2} &= l^2\left( q^2 + dt^2q^4 + dt^4q^6 + dt^6q^8 + dt^8q^{10} + dt^{10}q^{12} +  \dots \right)
    \;.
\end{align}
We see the formation of a derivative tower, germinating from the seed word, similar to the zero-coupling case.
However, there are two important differences.
Firstly, there is only one seed word at infinite coupling, as opposed to infinitely many seeds at zero coupling. This stems from the fact that only one super-multiplet contributes to the SLPF at infinite coupling~\eqref{eqn:SLPFStrong}, as opposed to the infinitely many that contribute at zero coupling~\eqref{eqn:zMultContent}.
Secondly, the number of derivatives grows by \emph{two} for every new element in the tower, as opposed to by one at zero coupling.
This is due to the fact that the derivative letters $\mathcal{D}_t$ in the zero coupling alphabet do not commute whereas those in the infinite coupling alphabet $\partial_t$ do, and therefore one needs to act with two derivatives $\partial_t$ to generate a two-particle word which is a conformal primary at infinite coupling.

Next, let us consider four-particle states for which we get
\begin{align}
    \label{eqn:4particleStrongMLPFdtdQ}
    \mathcal{Z}_{\infty}(q;l;dt;dQ)\bigg|_{l^4} &=
    l^4\left(q^4
    +2 {dt}^2 q^6
    +{dt}^3 q^7
    +4 {dt}^4 q^8
    +2 {dt}^5 q^9
    +6 {dt}^6 q^{10}+\dots\right)
    \;.
\end{align}
We note that
rather than a tower structure,
the seed four-particle state germinates 
a tree of derivatives.
We also see the appearance of a word with odd number of derivatives $\partial_t$ at order $q^7$.

Finally, let us look 
at higher-particle words 
for a given $\Delta_\infty$. 
The most interesting cases are from $\Delta_\infty = 8$ onwards.
For $\Delta_\infty = 8$ and $10$, we display the higher-particle words from equations~\eqref{eqn:strongSingletdQOq8} and~\eqref{eqn:strongSingletdQOq10}, and we generate them for $\Delta_\infty = 12$ and $14$\footnote{The MLPF also has words at $\mathcal{O}(q^7)$, $\mathcal{O}(q^9)$,  $\mathcal{O}(q^{11})$ and $\mathcal{O}(q^{13})$.}.
\begin{align}
    \label{eqn:strongSingletdQmaxq8}
    \mathcal{Z}_{\infty}(q;l;dt;dQ)\bigg|_{q^8;\,l^\texttt{max}}  
    &=
    q^8
    \bigg[
    l^8
    +l^6({\color{blue} dQ^4} +\dots)
    +\dots
    \bigg]
    \;,
    \\
    \label{eqn:strongSingletdQmaxq10}
    \begin{split}
    \mathcal{Z}_{\infty}(q;l;dt;dQ)\bigg|_{q^{10};\,l^\texttt{max}}  
    &=
    q^{10}
    \bigg[
    l^{10}
    +
    \dots
    +l^6(
    2 {\color{blue} dQ^4} {dt}^2
    +\dots
    )
    +\dots \bigg]
    \;,
    \end{split}
    \\
    \label{eqn:strongSingletdQmaxq12}
    \begin{split}
    \mathcal{Z}_{\infty}(q;l;dt;dQ)\bigg|_{q^{12};\,l^\texttt{max}}  
    &= q^{12}\bigg[
    l^{12}
    +
    \dots
    +
    l^8 \left({\color{red} dQ^8}
    +\dots
    \right) 
    +\dots
    +
    l^6 \left(5{\color{blue}dQ^4}dt^4\right)
    +\dots
    \bigg]
    \;,
    \end{split}
    \\
    \label{eqn:strongSingletdQmaxq14}
    \begin{split}
    \mathcal{Z}_{\infty}(q;l;dt;dQ)\bigg|_{q^{14};\,l^\texttt{max}}  
    &= q^{14}\bigg[
    l^{14}
    +
    \dots
    +
    l^8 \left(2{\color{red} dQ^8}dt^2
    +\dots
    \right) 
    +\dots
    +
    l^6 \left(12{\color{blue}dQ^4}dt^6\right)
    +\dots
    \bigg]
    \;.
    \end{split}
\end{align}
Let us first look at the highlighted terms at order $\left(q^{\Delta_\infty}l^{\frac{\Delta_\infty}{2}+2}\right)$
for $\Delta_\infty = 8,12,\dots$.
These are unique words whose fugacity content is exactly double of the super-primary states of the type described in equation~\eqref{eqn:SO3ChargedGroundState}.
Furthermore, these words form the seed of a derivative tree.
We have highlighted two trees in the expressions above: one starting at $\Delta_\infty = 8$ and another starting at $\Delta_\infty = 12$.

%% file: map.tex
\section{A Map Between Alphabets}
\label{sec:mapalphabets}

In this section, we will propose a simple map from the alphabet that captures the degrees of freedom at weak coupling, to that which captures the degrees of freedom at strong coupling. 
First, we will propose the map for single-particle states at weak coupling.
Then, we will minimally generalise this map to capture singlet two-particle states.
This will give us a testable prediction which we compare against the exact result obtained from integrability in the next section.

\subsection{Single-particle States}
\label{sec:SingleParticleMap}

\paragraph{Zero coupling.}
As we have reviewed in Section~\ref{sec:counting}, at zero coupling, the SLPF gets a contribution from an infinite number of super-multipletes, whose super-primaries have quantum numbers
$$
[0]_{1}^{(0,0)}\;,\;[2j]_{j+1}^{(0,0)}\;,\; j=0,1,\dots\;.
$$
Let us build all the single-particle super-primary states at zero coupling. 
As discussed in the Setup, from the field theory description, it is clear that the super-primaries of the $\mathcal{B}_1$- and $\mathcal{A}_2$-type with quantum numbers $[0]_1^{(0,1)}$ and $[0]_1^{(0,0)}$ are $\Phi_m$ and $\Phi_{||}$ respectively. 
The quantum numbers of super-primaries of the other supermultiplets also immediately suggest what could be their field content.
To illustrate this, let us 
consider the super-primary with quantum numbers
${[2]^{(0 ,0)}_{2}}$.
Three immediate possibilities for such an operator are $\mathcal{D}_i\Phi_{||}$, $\mathcal{F}_{ti}$ and $\epsilon_{ijk}\mathcal{F}_{jk}$. 
In reality, it can be shown~\cite{Cooke:2017qgm} that the true super-primary of the $\mathcal{A}_1$-type multiplet is
\begin{align}
    \Fpri_i \coloneqq
    \mathrm{i}\,\mathcal{F}_{ti} - \mathcal{D}_i \Phi_{||}\;.    
\end{align}
An orthogonal linear combination of these operators,
\begin{align}
    \disp_i \coloneqq
    \mathrm{i}\,\mathcal{F}_{ti} + \mathcal{D}_i \Phi_{||}\;,
\end{align}
can be shown~\cite{Cooke:2017qgm} to be a super-descendent primary of the $\mathcal{B}_1$-type multiplet with super-primary quantum number $[0]_1^{(0,1)}$. 
Finally, 
\begin{align}
    \Fdesc_i \coloneqq \epsilon_{ijk}\mathcal{F}_{jk}\;,
\end{align}
is a super-descendent~\cite{Cooke:2017qgm} of the $\mathcal{A}_2$-type multiplet with super-primary quantum number $[0]^{(0,0)}_{2}$.

More generally, suppose we want to construct a state with quantum numbers $[2j]^{(0,0)}_{j+1}$.
In order to do this, we need to create a combination of fields and derivatives, with $j$ $\mathrm{SO}(3)$ vector indices, and $\Delta_0 = j+1$.
The required vector indices may be generated in the following way:
\begin{itemize}
    \item Acting with $\mathcal{D}_i$ adds \emph{one} vector index to a potential state, and increases $\Delta_0$ by \emph{one}.
    \item The component of field strength $\mathcal{F}_{it}$ adds \emph{one} vector index to a potential state in increases $\Delta_0$ by \emph{two}.
    \item The component of field strength $\epsilon_{ikl}\mathcal{F}_{kl}$ adds \emph{one} vector index to a potential state in increases $\Delta_0$ by \emph{two}.
    \item A fermion bilinear contracted in such a way to transform as a singlet of $\mathrm{SO}(5)$ and a vector of $\mathrm{SO}(3)$ adds \emph{one} vector index to a potential state in increases $\Delta_0$ by \emph{three}.
\end{itemize}
Therefore, we have, for $j\geq 1$
\begin{xltabular}[c]{\textwidth}{C|C|C|C}
\text{Field content} &
 \Delta_0 &
 \mathrm{SO}(3)
 \text{ irrep} & \text{Allowed?}
 \\
 \midrule\midrule
 \rule{0pt}{3.5ex} 
 \mathcal{D}_{(i_{1}}\dots\mathcal{D}_{i_{j})}\Phi_{||}
 & j+1 & [2j] & 
 \text{\cmark}
 \\[1ex]
\hline
 \rule{0pt}{3.5ex}
 \mathcal{D}_{(i_{1}}\dots\mathcal{D}_{i_{j-1}}\mathcal{F}_{i_{j})t}
 & j+1 & [2j] & \text{\cmark} \\[1ex]
 \hline
 \rule{0pt}{3.5ex}
 \mathcal{D}_{(i_{1}}\dots\mathcal{D}_{i_{j-1}}\epsilon_{i_{j})kl}\mathcal{F}_{kl}
 & j+1 & [2j] & \text{\cmark} \\[1ex]
 \hline
 \rule{0pt}{3.5ex}
 \mathcal{D}_{(i_{1}}\dots\mathcal{D}_{i_{j-1}}{\Psi_{a\,\alpha}\Psi_{a\,\beta}}\sigma^{\alpha\beta}_{i_{j})}
 & j+2 & [2j] & 
 \text{\xmark}
 \\[1ex]
  \caption{
  Possibilities for a super-primary with quantum numbers $[2j]_{j+1}^{(0,0)}$\;.
  }
  \label{tab:buildSuperPrimaries}
\end{xltabular}
Clearly, 
only the combinations of fields and derivatives shown in the the first three rows of Table~\ref{tab:buildSuperPrimaries} above can generate such a state.
We conclude that to generate a state with the quantum numbers ${[2j]^{(0 ,0)}_{j+1}}$, one should take an appropriate number of derivatives $\mathcal{D}_i$, act with them on $\Fpri_i$, $\disp_i$ and $\Fdesc_i$,
and contract the derivatives in a way that they transform under the corresponding irrep of $\mathrm{SO}(3)$. 
One linear combination of the three generated operators will be a super-primary of the corresponding $\mathcal{A}_1$-type multiplet, and the other two will be super-descendents.

Let us introduce the notation 
\begin{align}
    \Fstar_{(i_1\dots i_j)} 
    \coloneqq 
    \mathcal{D}_{(i_1}D_{i_2}\dots \mathcal{D}_{i_{j-1}}
    \big[a_{j}\Fpri_{i_j)} + b_{j}\disp_{i_j)}
    +c_j \Fdesc_{i_j)}
\big]\;,
\end{align}
to denote the superprimary of the $\mathcal{A}_1$-type multiplet with superprimary quantum numbers $[2j]^{(0,0)}_{j+1}$, with $k\geq1$.
We conclude that $a_1 = 1$ and $b_1 = c_1 = 0$ from the preceding discussion.
There is no reason to a priori expect that $a_{j} = 1$ and $b_{j} = 0 = c_j$, for generic $j$, and as such, we will keep them undetermined.

\paragraph{Infinite coupling.}
It is an important question to ask as to where the single-particle states at weak coupling, end up at strong coupling.

Firstly, let us recall some known facts. 
The insertion $\Phi_{m}$ at weak coupling and the fluctuation $y_m$ at strong coupling are super-primaries of the same $\mathcal{B}_1$-type multiplet with protected scaling dimension equal to unity.
As such, it is clear that they map to each other.
It was speculated in~\cite{Giombi:2017cqn} and has since been explicitly shown in~\cite{Grabner:2020nis} that the insertion $\Phi_{||}$ maps to $y_m^2$.

At the time of writing this paper, there are no predictions as to where the other single-particle super-primary states at weak coupling end up at strong coupling.
We see from the analysis of equation~\eqref{eqn:SO3ChargedGroundState}, that these super-primary states, with quantum numbers $[2j]_{2(j+1)}^{(0,0)}$, are the natural candidates for the targets of a putative map from the super-primaries of the $\mathcal{A}_1$-type multiplets in the SLPF at zero coupling.

Let us give an example to illustrate the structure of these super-primary. For $j = 1$, we see from equation~\eqref{eqn:SO3ChargedGroundState} that have a three-particle state with $\Delta_\infty = 4$, containing the action of 2 supercharges. There are two candidates for such a state:
$$
f_i y_m^2\quad\text{and}\quad
\psi_{a\,\alpha}\psi_{b\,\beta}\sigma^{ab}_m\sigma_{i}^{\alpha\beta}y_m\;.
$$
The super-primary state is a linear combination of the above two.
In general we will have
5
operators in the basis, out of which we will construct the super-primary.
We introduce the notation $\fstar_{(i_1\dots i_j)}$ to denote the super-primary 
with quantum numbers $[2j]_{2(j+1)}^{(0,0)}$.

\paragraph{The map.}
For one-particle states at weak coupling, we propose the following map:
\begin{equation}
    \label{eqn:RuleSingleParticle}
    \begin{minipage}{\textwidth}
    \begin{displayquote} 
    {\it
    A single-particle state at weak coupling maps to the lowest lying state in the corresponding symmetry channel at strong coupling.}
    \end{displayquote}
    \end{minipage}
\end{equation}
In other words, for single-particle super-primary states, in addition to the known facts that
\begin{align}
    \label{eqn:MapPhim}
    \Phi_m &\to y_m\;,\\
    \label{eqn:MapPhipar}
    \Phi_{||}&\to y_m^2\;,
\end{align}
we have
\begin{align}
    \label{eqn:MapFstar}
    \Fstar_{(i_1\dots i_j)}\to \fstar_{(i_1\dots i_j)}\;.
\end{align}

\subsection{Two-particle Words}
\label{sec:2particleMap}

The next step in our analysis will be to consider two-particle words. 

\paragraph{Zero coupling.} 
The singlet MLPF restricted to two-particle words and refined to keep track of specific super-multiplet contribution is given in equation
~\eqref{eqn:2particleMLPFrefined}
and further refined to capture derivative and super-descendent contribution in equation
~\eqref{eqn:2particleMLPFrefineddt}.

From a closer look at the quantum numbers of the super-primaries of the various supermultiplets contributing to the SLPF at weak coupling~\eqref{eqn:zMultContent}, we see that  
for $j\geq 1$, the multiplet associated with the fugacity $l_{2+j}$ has a super-primary with quantum numbers 
${[2j]^{(0 ,0)}_{j+1}}$. 
Clearly one can make a singlet word by taking two such states and contracting the $\mathrm{SO}(3)$ indices. 
From the lack of 
$dQ$ fugacities in the refined two-particle MLPF~\eqref{eqn:2particleMLPFrefineddt},
and the fact that its dimension is $\Delta_0 = 2j+2$,
we can conclude that this word contributes to the super-primary state.

With this picture in mind, let us try to understand what could be the possible singlet two-particle words that can be built at a given bare dimension, which can without loss of generality be taken to be even. Let $\Delta_0 = 2 k$. We are looking for words with quantum numbers $[0]_{2k}^{(0,0)}$. From the counting formula~\eqref{eqn:2particleMLPFrefined}, we expect there to be $k+1$ such words,
these are
$$
[\Phi_m^2]_{2k-2}\;,\quad [\Phi_{||}^2]_{2k-2}\;,\quad [(
\Fpri_{i}
)^2]_{2k-4}\;,\quad [(\Fstar_{(i_1i_2)})^2]_{2k-6}\;,
\dots\;,\;
(\Fstar_{(i_1\dots i_k)})^2\;.
$$
Here the notation $[\dots]_n$ 
means that we act $n$-times with the covariant derivative $\mathcal{D}_t$.

\paragraph{Infinite coupling.}

From equation~\eqref{eqn:2particleStrongMLPFdtdQ}, we see that there is one tower of two-particle words germinated by a neutral seed state with $\Delta_\infty = 2$.
The only possibility for this seed state is $y_m^2$, and therefore we identify the two-particle tower as
    $$y_m^2\;,\;
    [y_m^2]_{2}\;,
    [y_m^2]_{4}\;,
    \dots\;.$$
From equation~\eqref{eqn:4particleStrongMLPFdtdQ} we see that the derivative tree of four-particle words is generated by a neutral state with $\Delta_\infty = 4$. We identify this word to be $y_m^2y_n^2$.

Next, recall that we had identified in equations~\eqref{eqn:strongSingletdQmaxq8} and~\eqref{eqn:strongSingletdQmaxq12}, words which had the double the fugacity content as the super-primary states $\fstar_{(i_i\dots i_j)}$ described in equation~\eqref{eqn:SO3ChargedGroundState}.
While it is tempting to interpret these words as the ``squares'' of the states $\fstar_{(i_i\dots i_j)}$, with the $\mathrm{SO}(3)$ indices contracted to create a singlet, it is not possible to say for sure that this is the case.
In principle, there could be other combination of degrees of freedom that give the same fugacity content.  
This is in contrast to the zero coupling case, where, due to the fact that we have specific letters for each $\Fstar_{(i_1\dots i_j)}$, it is possible to say for sure that the two-particle seed singlet words are of the form $(\Fstar_{(i_1\dots i_j)})^2$.
We adopt 
the notation 
$\ftilde{2},\ftilde{3}, \dots$ 
to describe the seed words from equations~\eqref{eqn:strongSingletdQmaxq8} and~\eqref{eqn:strongSingletdQmaxq12}.

\paragraph{The map.}
Let us 
now try to extend the map we proposed in the Section~\ref{sec:SingleParticleMap} to two-particle words at weak coupling\footnote{We are extremely grateful to Barak Gabai, Victor Gorbenko, Bendeguz Offertaler and Miguel Paulos for detailed discussion about the passages below.}.

Let us proceed step by step. 
Let us start with the two-particle tower $[\Phi_m^2]_{n}$. At each value of $n$, this tower generates one operator with $\Delta_0 = 2+n$.
Let $n=0$, we then we have $\Phi_m^2$. 
Based on the single-particle map~\eqref{eqn:MapPhim} of $\Phi_m\to y_m$, 
a first attempt would be to map this word to the seed state of the $[y_m^2]_{2n}$ tower at strong coupling, \textit{i.e.} to $y_m^2$. 
However we already know from~\cite{Giombi:2017cqn,Grabner:2020nis} (\textit{cf.} equation~\eqref{eqn:MapPhipar}), that this state maps to $\Phi_{||}$.
The second attempt would be then to map this word to the next element in the tower: $[y_m^2]_2$.
The next word in the weak coupling tower, $[\Phi_m^2]_1$ then gets mapped to the next available level in the strong coupling tower, which is $[y_m^2]_4$.
In this way, one can map the entire weak coupling tower to the entire strong coupling tower.
Due to the initial shift in dimensions due to the already occupied level at $y_m^2$, 
we get the map
\begin{align}
    [\Phi_{m}^{2}]_n \rightarrow [y_m^2]_{2+2n}\;.
\end{align}
This also implies that we map the derivatives as
\begin{align}
    \mathcal{D}_t\to\partial_t^2\;.
\end{align}
We will indeed do so in the passages that follow.
Next, let us consider the tower $[\Phi_{||}^2]_n$.
Based on the single-particle map~\eqref{eqn:MapPhipar} of $\Phi_{||}$ to $y_m^2$, we propose that the entire two particle tower maps as
\begin{align}
    [\Phi_{||}^{2}]_n \rightarrow [y_m^2y_k^2]_{2n}\;.
\end{align}
Here, $[y_m^2y_k^2]_{2n}$ represents \emph{one} 
one of the various words with $2n$-derivatives in the 
derivative tree generated by $y_m^2y_k^2$ in equation~\eqref{eqn:4particleStrongMLPFdtdQ}.

Finally, let us consider towers of the form $[(\Fstar_{(i_1\dots i_j)})^2]_n$. 
Based on the single particle map~\eqref{eqn:MapFstar} of $\Fstar_{(i_1\dots i_j)}\to \fstar_{(i_1\dots i_j)}$, 
due to the fact that $\ftilde{j}$ has double the fugacity content of $\fstar_{(i_1\dots i_j)}$,
we propose
\begin{align}
    [(\Fstar_{(i_1\dots i_j)})^2]_n \rightarrow [\ftilde{j}]_{2n}\;.
\end{align}
where again the notation on the RHS represents \emph{one} 
one of the various words with $2n$-derivatives in the 
derivative tree generated by $\ftilde{j}$.

Therefore, for all two-particle singlet words at $\Delta_0 = 2 k$, we get
\begin{align}\label{eqn:2particlemap}
    \begin{split}
        [\Phi_m^2]_{2k-2} &\to [y_m^2]_{4k-2} \;,\\
        [\Phi_{||}^2]_{2k-2} &\to [y_m^2y_n^2]_{4k-4}\;,\\
        [(\Fpri_{i})^2]_{2k-4} &\to [\ftilde{1}]_{4k-8} \;,\\
        [(\Fstar_{(i_1i_2)})^2]_{2k-6} &\to [\ftilde{2}]_{4k-10}\;,\\
        &\dots\\
        (\Fstar_{(i_1\dots i_k)})^2 &\to \ftilde{k} \;.
    \end{split}
\end{align}
Notice that under this map, the images of all two-particle words with $\Delta_0 = 2k$, are words with $\Delta_\infty = 4k$.

The true super-primary states at some $\Delta_0$ can in principle contain contributions from words made up of any number of particles. 
Therefore, 
strictly speaking 
our map should be viewed only as a 
map from a set of words at zero coupling to a set of words at strong coupling. 
If indeed the super-primary can be built just out of two-particle words, then our map \emph{predicts} 
that the dimensions of these two-particle singlet super-primary states \emph{double} as they evolve from zero to infinite coupling.
This is a concrete prediction, that we will test in the next section, by comparing with the exact spectrum of two-particle states, obtained using integrability methods.

%% file: qsc_v2.tex
\section{Comparison to Exact Spectrum}
\label{sec:integrability}

The non-perturbative spectrum of insertions in the 1D dSCFT
can be constructed using integrability-based methods.
One of the most powerful such methods goes by the name of Quantum Spectral Curve (QSC) and it has been used to solve the spectral problem exactly local operators in planar $\mathcal{N} = 4$ SYM~\cite{Gromov:2014caa,Gromov:2015wca, Gromov:2015dfa,Gromov:2023hzc}.
The low-lying spectrum spectrum of insertions in the 1D dSCFT
was obtained using the QSC~\cite{ Grabner:2020nis,Julius:2021uka,Cavaglia:2021bnz,Cavaglia:2024dkk}, which enables us to track the states from weak to strong coupling. 
In the following, we will briefly review and extend the QSC for the 1D dSCFT.
Then, we compare the predictions of the map from Section~\ref{sec:mapalphabets} to the exact numerical spectrum obtained from the QSC.

\paragraph{Quantum Spectral Curve.} 
The QSC for the spectrum in the 1D dSCFT has already been extensively developed.
In particular, the current understanding~\cite{ Grabner:2020nis,Julius:2021uka,Cavaglia:2021bnz,Cavaglia:2024dkk} covers all states in the theory which are neutral under $\mathrm{SO}(3)$ but have arbitrary charges under $\mathrm{Sp}(4)$, at least in principle. 
For the purposes of this paper, we 
need to tackle some sectors that are charged under $\mathrm{SO}(3)$.
We propose a straightforward generalisation to capture these 
sectors. 

The QSC for insertions in the 1D dSCFT coincides with the QSC for local operators in planar $\mathcal{N} = 4$ SYM at one-loop. 
This is because, at one-loop, one can use the method of images to identify the open spin-chain description of the 1D dSCFT (with the supersymmetric Wilson line providing integrable boundary conditions) to the closed spin-chain description of $\mathcal{N} = 4$ SYM~\cite{Cavaglia:2021bnz} (see also~\cite{Gromov:2021ahm}).
States in planar $\mathcal{N} = 4$ SYM can be identified using their \texttt{State ID}~\cite{Gromov:2023hzc}:
\begin{align}
    \texttt{State ID}^{(4\text{D})} :
    {}_{\Delta^{(4\text{D})}_0}[n_{\textbf{b}_1}\;n_{\textbf{b}_2}\;n_{\textbf{f}_1}\;n_{\textbf{f}_2}\;n_{\textbf{f}_3}\;n_{\textbf{f}_4}\;n_{\textbf{a}_1}\;n_{\textbf{a}_2}]_{\texttt{sol}^{(4\text{D})}}\;,
\end{align}
where $n_{{\bf a}_i},n_{{\bf b}_i},n_{{\bf f}_i}$ are oscillator numbers~\cite{Marboe:2017dmb}, $\texttt{sol}^{(4\text{D})}$ is a multiplicity label and $\Delta^{(4\text{D})}_0$ is the bare dimension of the state in $\mathcal{N} = 4$ SYM.

In general, there is a multiplicity of states with the same dimension $\Delta_0$ in a particular global charged sector at zero coupling in the 1D dSCFT.
The multiplicities for the singlet sector are explicitly given in equation~\eqref{countingweak}.
These multiplicities are also captured using a \texttt{State ID}~\cite{Cavaglia:2024dkk}:
\begin{align}
     \texttt{State ID}^{(1\text{D})} :
     {}_{\Delta_0}[2j\;;\;m\;n\;;T]_{\texttt{sol}}
    \;.
\end{align}
Here, apart from the quantum numbers $[j]$ and $(m,n)$ capturing the global symmetries $\mathrm{SO}(3)$ and $\mathrm{Sp}(4)$, we also introduce two multiplicity labels $T$, called the twist, and $\texttt{sol}$.
Generalising the existing prescription~\cite{Cavaglia:2021bnz,Cavaglia:2024dkk} of $\texttt{State ID}^{(1\text{D})}\to\texttt{State ID}^{(4\text{D})}$, which captures the $j=0$ sectors, we propose to assign
\begin{align}
    \label{eqn:map1D4D}
    \begin{split}
        \Delta_0^{(4\text{D})} &\to 2\Delta_0+1\;,\\
        n_{\textbf{b}_1} &\to (\Delta_{0}-T)\;, \\
        n_{\textbf{b}_2} &\to (\Delta_{0} + 2j - T)\;, \\
        n_{\textbf{f}_1} &\to (T + m + n + 1 + j)\;, \\
        n_{\textbf{f}_2} &\to (T + n + 1 - j)\;, \\
        n_{\textbf{f}_3} &\to (T - n - j)\;, \\
        n_{\textbf{f}_4} &\to (T - m - n -j)\;, \\
        n_{\textbf{a}_1} &\to (\Delta_{0} + 2j - T)\;, \\
        n_{\textbf{a}_2} &\to (\Delta_{0}-T)\;. \\
    \end{split}
\end{align}
For a given choice of $[j]$, $(m,n)$ and $\Delta_0$, and all possible choices of twist $T$, we find all the states in planar $\mathcal{N} = 4$ SYM, that have compatible oscillator numbers.
A subset of these states, which respect the integrable boundary conditions imposed by the supersymmetric Wilson line, are the ones that we are after.

Below, the overall QSC procedure is briefly discussed.
The computation follows the same canvas as~\cite{Gromov:2014caa, Gromov:2015wca}. 
We start with a distinguished set of eight functions of a complex variable $u$ called the spectral parameter.
These functions are denoted as 
$\mathbf{P}_{a}$ and $\mathbf{Q}_i$ with $a,i = 1,\dots,4$.
They have a specific branch-cut structure in the complex $u$-plane.
In particular, the $\mathbf{P}_a$ have only one branch cut, stretching between $-2g$ and $+2g$, where $g \coloneqq \sqrt{\lambda}/(4\pi)$ is related to the 't Hooft coupling. 
This allows them to be parameterised as
\begin{equation}
    \mathbf{P}_a(u) = \mathbb{A}_{a}\,u^{\texttt{powP}_{a}} \left(1 + \sum_{n=1}^{\infty} \frac{c_{a,n}}{x^{n}} \right), 
\end{equation}
with $x(u) = \frac{u + \sqrt{u-2g}\sqrt{u+2g}}{2g}$ being the Zhukovsky variable. 
The large-$u$ asymptotics coefficients $\texttt{powP}_a$ and $\mathbb{A}_a$ are completely described by global charges of a state, we have
\begin{align}
    \mathbf{P}_a &\sim \mathbb{A}_a\,u^{\mathtt{powP}_{a}}\;, \\
    \mathtt{powP}_a &= {\left\{-m-n-\frac{5}{2},-n-\frac{3}{2},n+\frac{1}{2},m+n+\frac{3}{2}\right\}}_{a}\;,\\
    \nonumber
    \mathbb{A}_a &= \left\{1,1,
    \right.\\
    \nonumber
    &\left.
    -\frac{i (-\Delta +j+n) (\Delta +j- n+1) (j- (\Delta +n+2)) (j+
   (\Delta +n+3))}{2 (m+1) (n+1) (m+2 n+3)}
    ,
   \right.\\
    &\left.
   \nonumber
   -
   \frac{1}{2 (m+1) (m+n+2) (m+2
   n+3)}  
   \bigg[i (j+ (\Delta -m-n)) (j- (\Delta
   -m-n-1)) 
   \right.\\
    &\left.
   (j- (\Delta +m+n+3)) (j+ (\Delta +m+n+4))\bigg]
   \right\}_a
    \;.
\end{align}
The $\mathbf{Q}_i$ on the other hand have a complicated branch cut structure in the complex-$u$ plane. However, their large-$u$ asymptotics are also governed by the global charges:
\begin{align}
\mathbf{Q}_i &\sim \mathbb{B}_{i}\, u^{\texttt{powQ}_{i}}\\
\mathtt{powQ}_i &= 
    \left\{\Delta +{j}+\frac{3}{2},\Delta -{j}+\frac{1}{2},-\Delta
   +{j}-\frac{3}{2},-\Delta -{j}-\frac{5}{2}\right\}_i
    \;.
    \nonumber
    \\
    \mathbb{B}_i &= \left\{
    \frac{i (\Delta +j-n+1) (j+ (\Delta +n+3)) (j- (-\Delta +m+n)) (j+ (\Delta
   +m+n+4))}{2 (2 \Delta +3) (2j+1) (\Delta +j+2)}
   \right.
   \nonumber
   \\
   \nonumber
    &\left.
   -\frac{i (-\Delta +j+ n) (j-
   (\Delta +n+2)) (j+ (-\Delta +m+n+1)) (j- (\Delta +m+n+3))}{2 (2 \Delta +3) (2j+1)
   (j- (\Delta +1))},
    \right.\\
    &\left.
    1,1
   \right\}_i
    \;.
\end{align}
The $\mathbf{P}_a$ and $\mathbf{Q}_i$ satisfy a fourth order finite-difference Baxter equation~\cite{Alfimov:2014bwa}. 
This equation can be used to obtain the $c_{a,n}$ coefficients describing the $\mathbf{P}_a$ at tree-level.
This is fed into an iterative procedure that systematically produces the weak coupling solution for the $\mathbf{P}_a$ and $\mathbf{Q}_i$~\cite{Gromov:2015vua,1DSolver} by
incorporating the discontinuity relation on the $\mathbf{Q}_i$~\cite{Gromov:2015dfa,Grabner:2020nis}.
The perturbative solution serves as a 
as starting point of the 
non-perturbative 
numerical solution~\cite{Grabner:2020nis} for the scaling dimension $\Delta(\lambda)$.

\paragraph{Single-particle spectrum.}

Firstly, let us test the map~\eqref{eqn:RuleSingleParticle} on single-particle states given in Section~\ref{sec:SingleParticleMap}.
For that, let us display the plot of non-perturbative scaling dimensions for 
the lowest lying operators with quantum numbers $[2j]_{j+1}^{(0,0)}$. 
The state at $j=0$ it the super-primary with quantum numbers $[0]_{1}^{(0,0)}$  represented by 
$\Phi_{||}$ at zero coupling, whose dimension was obtained in~\cite{Grabner:2020nis}.
The states with $j=1,2,3$ are the super-primaries with quantum numbers $[2]_{2}^{(0,0)},[4]_{3}^{(0,0)}$ and $[6]_{4}^{(0,0)}$  represented by 
$\Fpri_{i},\Fstar_{(i j)}$ and $\Fstar_{(i j k)}$ at zero coupling, and their dimensions are  new results.
We also display the corresponding anomalous dimensions up to two loops at weak coupling analytically. 
The perturbative dimension for the state with quantum numbers $[0]_{1}^{0,0}$ was known previously~\cite{Alday:2007he,Bruser:2018jnc,Grabner:2020nis}, whereas the perturbative results for the other states are new.
We have
\begin{align}
    [0]_{0}^{(0,0)}: \Delta &= 1 + 4 g^2 - 16 g^4 + \mathrm{O}(g^6)\;,\\
    [2]_{2}^{(0,0)}: \Delta &= 2 + 6 g^2 - 24 g^4 + \mathrm{O}(g^6)\;,\\
    [4]_{3}^{(0,0)}: \Delta &= 3 + \frac{22}{3} g^2 - \frac{790}{27} g^4 + \mathrm{O}(g^6)\;,\\
    [6]_{4}^{(0,0)}: \Delta &= 4 + \frac{25}{3} g^2 - \frac{895}{27} g^4 + \mathrm{O}(g^6)\;.
\end{align}
\begin{figure}[htbp]
  \centering
  \includegraphics[width=0.8\linewidth]{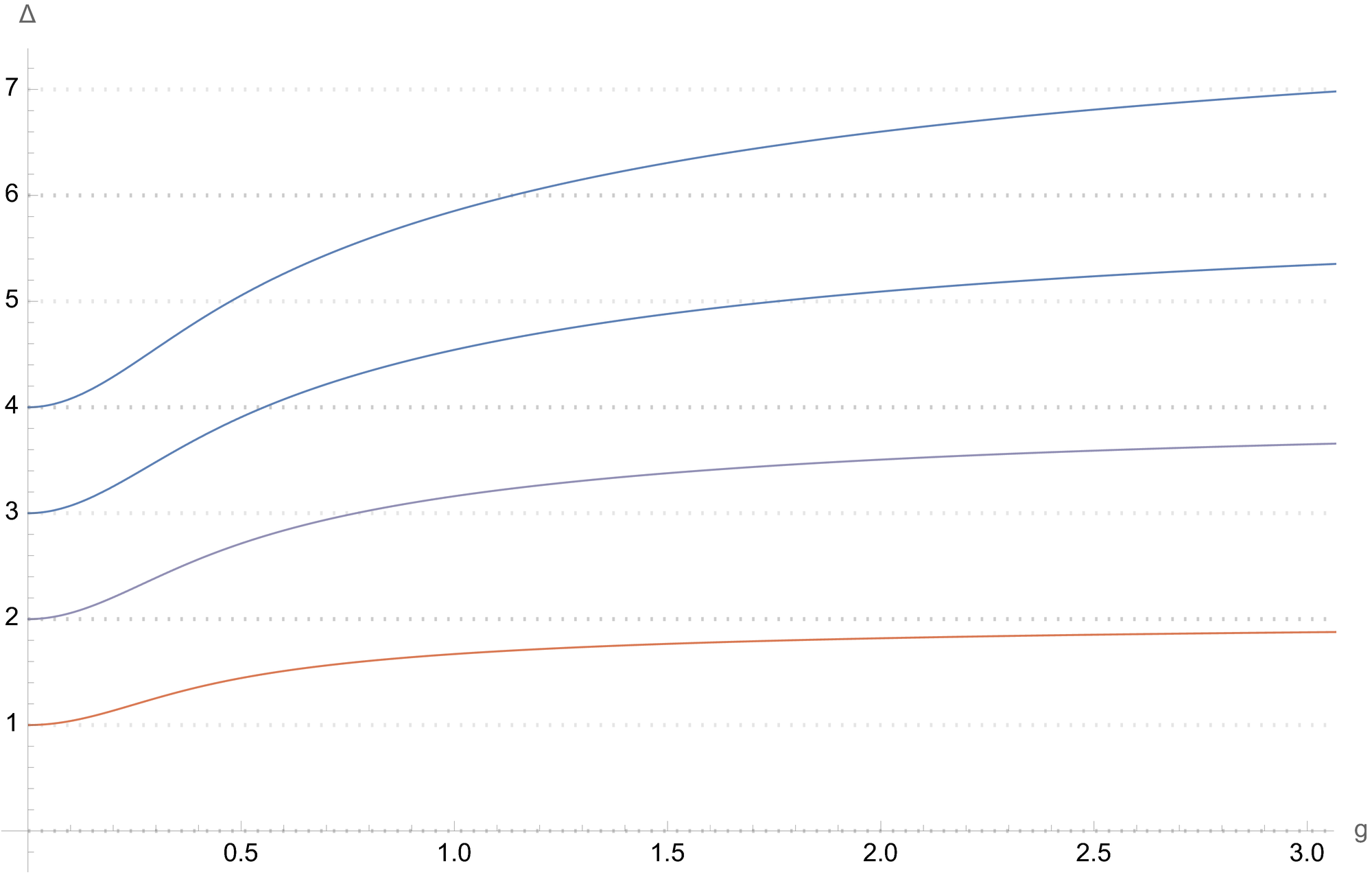}
  \caption{Single-particle spectrum. 
  The operators are the super-primaries of the $[2j]_{j+1}^{(0,0)}$ supermultiplets, whose dimension goes from $\Delta_0=j+1$ to $\Delta_{\infty}=2(j+1)$. 
  We display states with $j=0,\dots,3$.
  }
  \label{fig:single-particle}
\end{figure}
We see from 
Figure \ref{fig:single-particle}, that all four examples have $\Delta_{\infty} = 2 \Delta_{0}$, consistent with the prediction~\eqref{eqn:RuleSingleParticle}. 

\begin{figure}[h]
  \centering
  \includegraphics[width=0.9\linewidth]{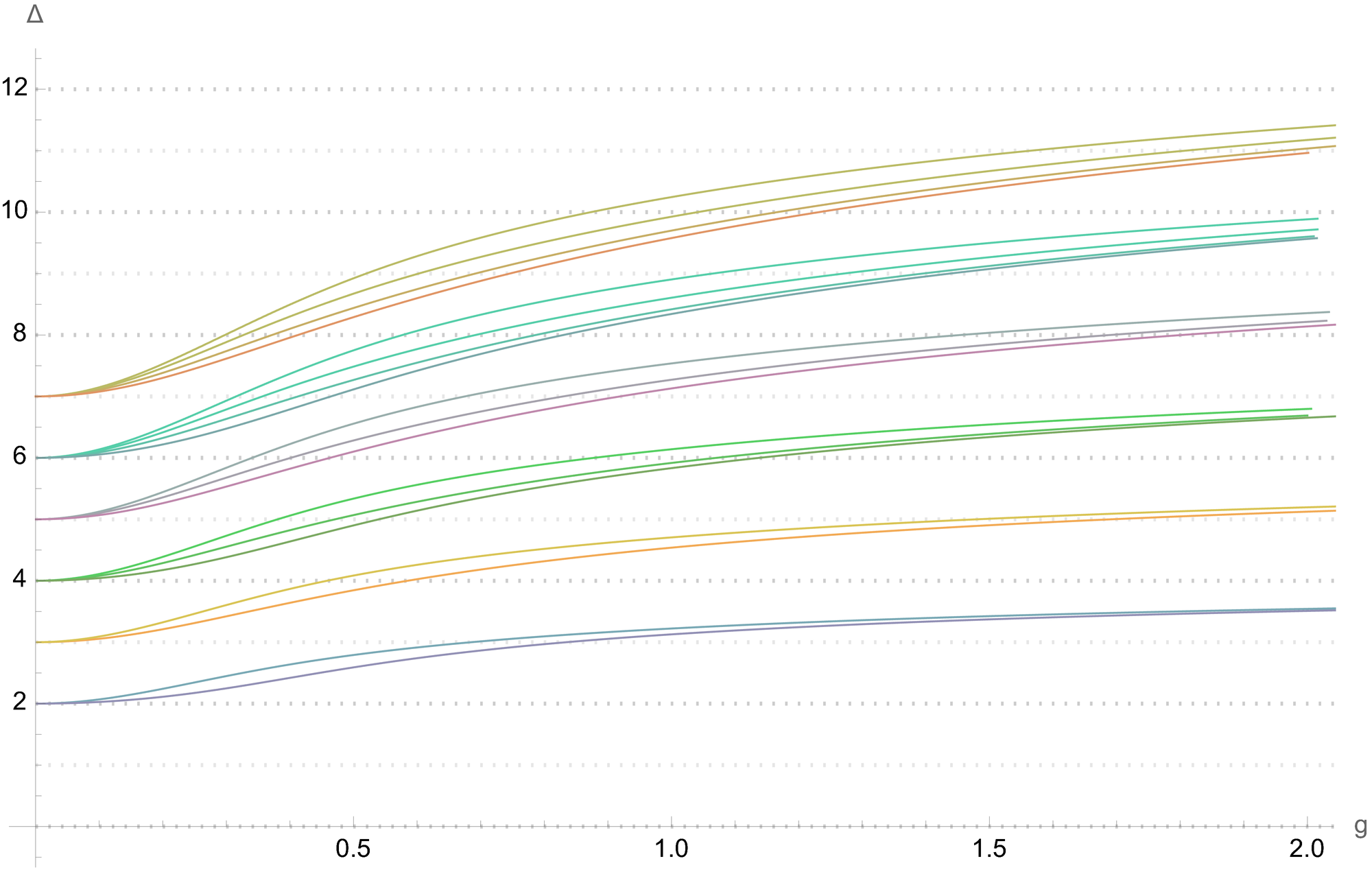}
  \caption{Twist-2 spectrum of singlet operators in $[0]_{\Delta}^{(0,0)}$ multiplets. We suppose that this spectrum is equivalent to the two-particle spectrum.}
  \label{fig:TwoParticleSinglet}
\end{figure}

\paragraph{Two-particle spectrum.} 
We believe that, at least at one-loop, the 
two-particle words form a basis of two-particle states, and these states correspond to those with twist $T=2$ in the QSC.
Let us describe below what this is based on.

First of all, twist-2 solutions of the Baxter equation grow as $1 + \texttt{floor}\left(\frac{\Delta_0}{2}\right)$ with $\Delta_0$~\cite{Julius:2021uka}, which is precisely the growth of two-particle words \eqref{eqn:2particleMLPFrefined}.
We have explicitly checked this growth up to $\Delta_0=10$.

At this point in the conversation, let us recall that 
for states with composed purely of scalars, 
the spin-chain Hamiltonian that gives the anomalous dimensions and eigenstates of the Dilatation operator at one-loop is known~\cite{Correa:2018fgz}.
The eigenstates can be written explicitly in terms of fields.
The one-loop low-lying singlet spectrum is presented in Table \ref{tab:HamiltonianDiag}. 
\begin{table}[ht!]
\centering
\small
\begin{tabular}{C|C|C|C|C}
\hline
     \Delta_0  & \#\,g^2 & \Delta_{\infty} & \text{Wavefunction for highest twist states} & T \\[2ex] 
    \hline\hline 
     1 & 4 & 2 & \Phi_{||} & 1 \\[2ex]\hline
    2 & 2.764 & 4 & \Phi_{||}^2 - \frac{1}{\sqrt{5}}\Phi_m\Phi_m & 2 \\[2ex]\hline
    2 & 7.236 & 4 & \Phi_{||}^2 + \frac{1}{\sqrt{5}}\Phi_m\Phi_m & 2 \\[2ex]\hline
   3 & 2.132 & 6 & \Phi_{||}^3 + a_{3,1;1}\,(\Phi_{||}\,\Phi_m\Phi_m + \Phi_m\Phi_m\Phi_{||}) + a_{3,1;2}\,\Phi_m\Phi_{||}\,\Phi_m & 3 \\[2ex]\hline
     3 & 5.103 & 6 & \Phi_{||}^3 + a_{3,2;1}\,(\Phi_{||}\,\Phi_m\,\Phi_m + \Phi_m\,\Phi_m\,\Phi_{||}) + a_{3,2;2}\,\Phi_m\,\Phi_{||}\,\Phi_m & 3 \\[2ex]\hline
    3 & \frac{1}{3}(23 - \sqrt{37}) & 6 &  & 2 \\[2ex]\hline
    3 & 9 & 7 & \Phi_{||}\,\Phi_m\,\Phi_m - \Phi_m\,\Phi_m\,\Phi_{||} & 3 \\[2ex]\hline
    3 & \frac{1}{3}(23 + \sqrt{37}) & 6 &  & 2 \\[2ex]\hline
    3 & 11.765 & 8 & \Phi_{||}^3 + a_{3,4;1}\,(\Phi_{||}\,\Phi_m\,\Phi_m + \Phi_m\,\Phi_m\,\Phi_{||}) + a_{3,4;2}\,\Phi_m\,\Phi_{||}\,\Phi_m & 3 \\[2ex]\hline
\end{tabular}
\captionof{table}{
\label{tab:HamiltonianDiag} 
The one-loop low-lying singlet spectrum.
Explicit eigenstates in terms of field content is given for highest twist states. 
For $\Delta_0 = 2,3$, the twist-2 states ($T=2$) correspond to two-particle states.} 
\end{table}
For $\Delta_0=2$, there are two singlet eigenstates of the Hamiltonian.
Both these states are composed of two-particle words, as there are no higher-particle words possible with $\Delta_0=2$.
For $\Delta_0 = 3$, 
there are four singlet eigenstates of the Hamiltonian.
All four of them are composed purely of three-particle words as can be seen from Table~\ref{tab:HamiltonianDiag}.
Their anomalous dimension corresponds to the $T=3$ solutions of the Baxter equation. 
The solutions of $T=2$ Baxter equation gives the anomalous dimensions of the two other states, and we conclude that these are two-particle states.

Lastly, the fact that we can map states in the 1D dSCFT back to those in $\mathcal{N}=4$ SYM at one-loop using the map~\eqref{eqn:map1D4D} can be used to argue that states in the 1D dSCFT with given quantum numbers and twist can only mix among themselves at one-loop order. 
This is the case for their counterparts in $\mathcal{N}=4$ SYM~\cite{Marboe:2017dmb}. 
Of course, one cannot rule out the possibility of mixing at higher-loop orders, as is the case in~$\mathcal{N}=4$ SYM.

The three arguments described above, lead us to the assumption that the notion of two-particle states in the 1D dSCFT is meaningful, at least at the one-loop order, and that the 
twist-2 solutions of the Baxter equation (and correspondingly of the QSC), are precisely these states.
It remains to check if the prediction from Section~\ref{sec:2particleMap} holds on these states.

The twist-2 spectrum 
was already obtained using the QSC in~\cite{Cavaglia:2021bnz,Julius:2021uka}, and it is
displayed in Figure~\ref{fig:TwoParticleSinglet}. 
We observe, that indeed, the dimension of the twist-2 (two-particle) states doubles from zero to infinite coupling, \textit{i.e.} $\Delta_0 \to \Delta_\infty = 2 \Delta_{0}$!
This confirms the prediction from Section~\ref{sec:2particleMap}, and gives credence to our smooth map of two-particle words.

%% file: pade_v2.tex
\section{An Application: Pad\'e Approximation}
\label{sec:pade}

In this section, the Pad\'e approximation for the spectrum is tested in comparison with the exact spectrum available from integrability. We test how good is the approximation compared to the very precise integrability-based data, for the first low-lying states. For those states, some perturbative orders at weak and strong coupling are known: these are the orders that constrain the Pad\'e for us.

The motivation of this study is to see if the Pad\'e method can be applied for higher states, if some perturbative results become known for them, even in case the exact non-perturbative spectral data has not been obtained. 
Armed with the map between two-particle states, one has grounds for using the two-sided Pad\'e, which could give some data for the spectrum at intermediate coupling.

We perform the Pad\'e analysis in terms of the a new coupling $y(g)$ defined as 
\begin{equation}
    \Delta_1 = 1 + y(g)\;,
\end{equation}
This is the anomalous dimension of the first two-particle singlet state $\Delta_{1}$. 
The reason for choosing this coupling is two-fold. Firstly, the anomalous dimension of the lowest lying operator functions as natural scheme independent interpolant, and therefore this choice will also make it easier to interpret and generalise the lessons learnt in this setup, to other contexts. 
Secondly, it is also a more convenient choice, as it takes values: $y \in [0,1]$, as opposed to the 't Hooft coupling, which goes from 0 to $\infty$.
The Pad\'e approximant is parameterised as
\begin{align}\label{eqn:PadeFunc}
    \texttt{Pad\'e}(y) = \frac{\Delta_0 +
    \sum_{k=1}^{M} \mathfrak{a}_ky^k}{1+\sum_{l=1}^N\mathfrak{b}_ly^l}\;,
\end{align}
The limits of the sums $M$ and $N$ and the parameters $\mathfrak{a}_k$ and $\mathfrak{b}_l$ are appropriately fixed by demanding a match with perturbative expansions up to some given order.
There will usually be many possible choices of $M$ and $N$ for a given state. 
We choose the best values by seeing which approximant differs the least from the exact result. 
In doing so, we observe some general patterns that may function as useful thumb-rules for generating Pad\'e approximations in the cases where the exact result is not known.


Below, we present our Pad\'e analysis. We perform a two-sided Pad\'e approximation and always input three sub-leading orders at weak coupling. 
We input one or two sub-leading orders at strong coupling, depending on what is currently known in the literature.
\begin{figure}[h!]
  \centering
  \includegraphics[width=0.8\textwidth]{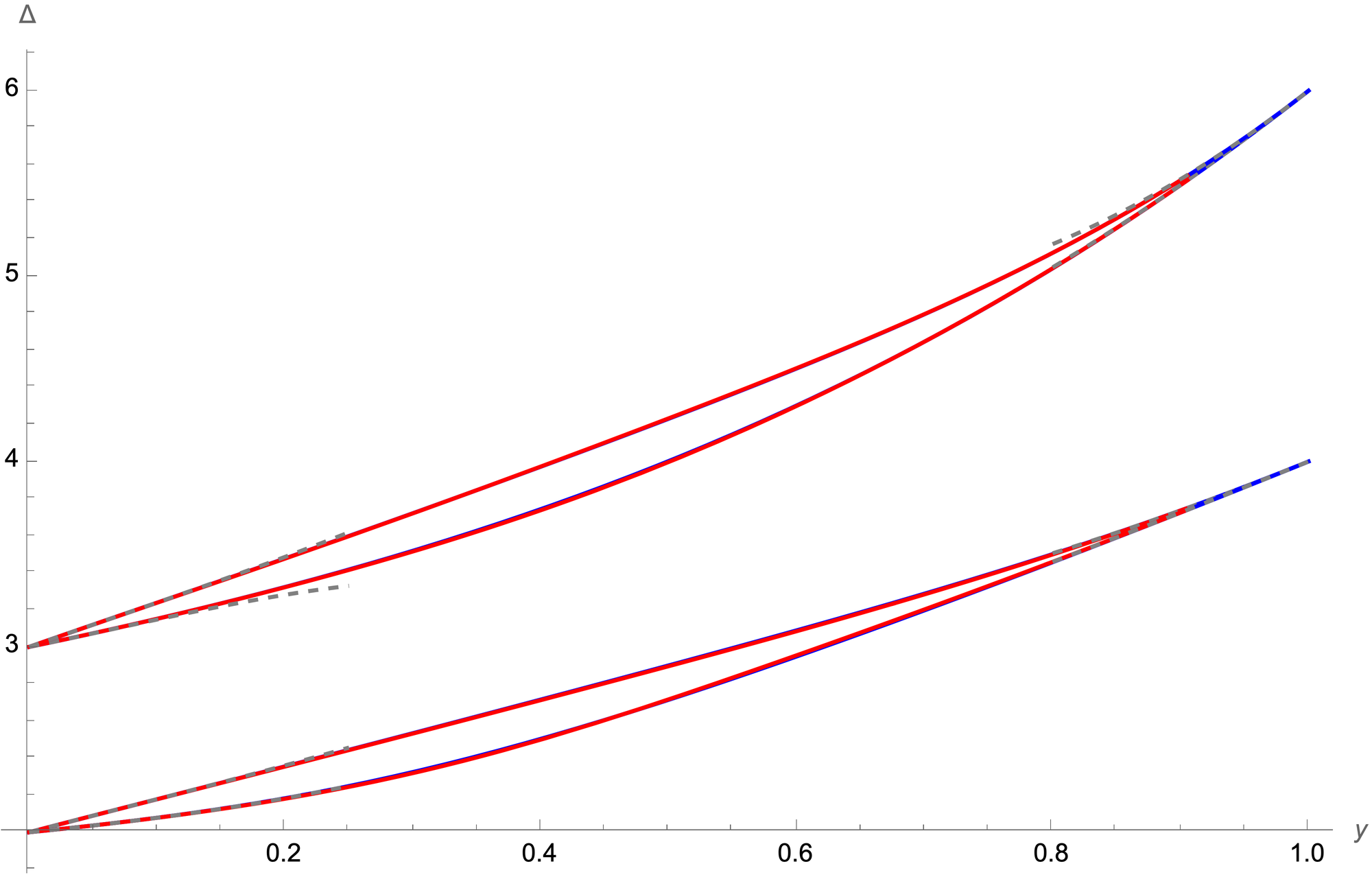}
  \caption{Spectrum of $\Delta_{2,1}$, $\Delta_{2,2}$, $\Delta_{3,1}$ and $\Delta_{3,2}$ operators displayed together with the best  respective Pad\'e approximants as per Table~\ref{tab:2ptPadeBestDel23}: the exact integrability data is shown in red, and the approximation in blue. The weak and strong perturbative data is shown with the dashed lines. We display the approximant with $M=4$ and $N=2$ for $\Delta_{3,1}$.}
  \label{fig:Pade}
\end{figure}

\begin{figure}[h!]
  \centering
  \begin{subfigure}{0.45\textwidth}
    \centering
    \includegraphics[width=\textwidth]{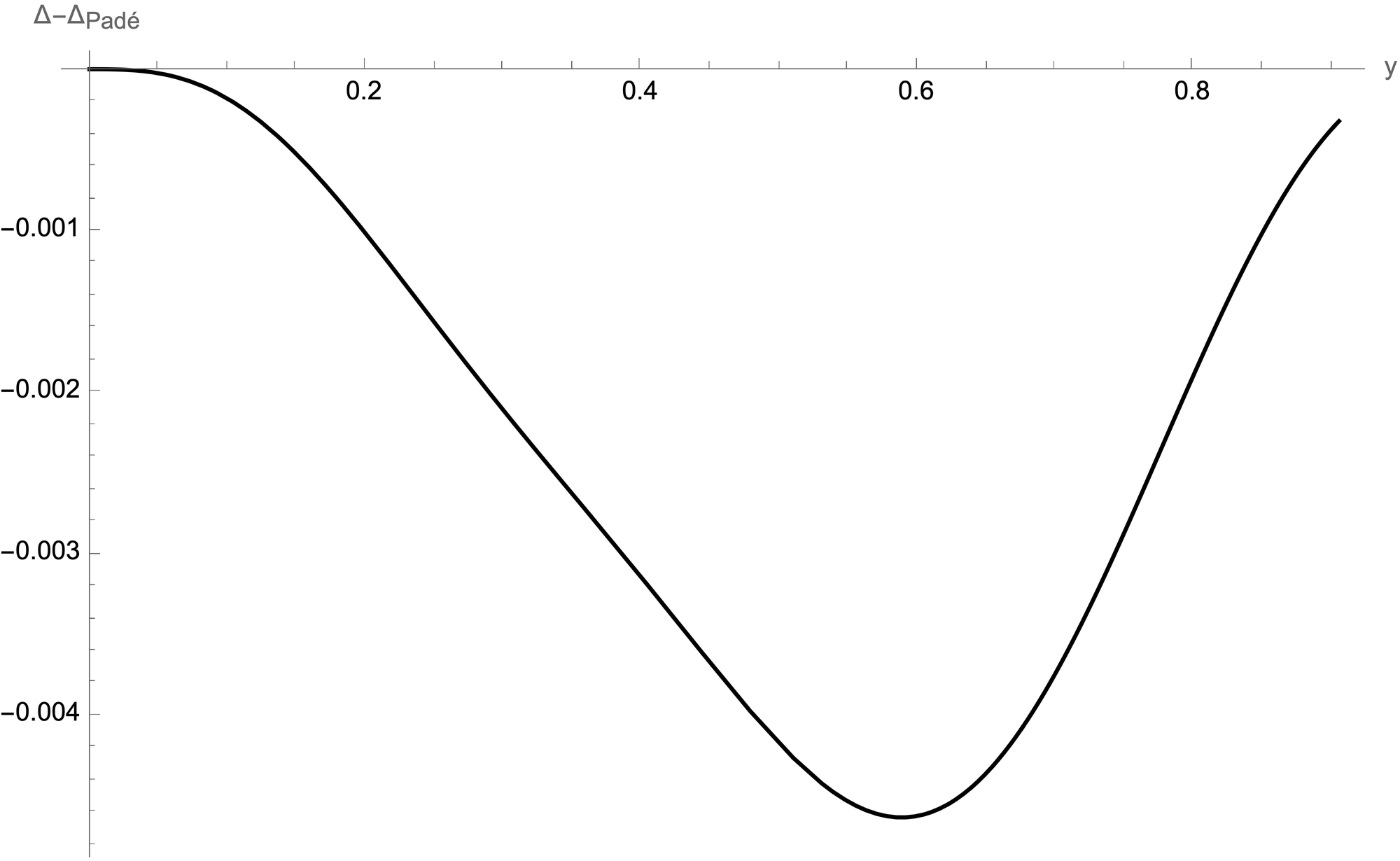}
    \caption{Error for $\Delta_{2,1}$ with $M=6$ and $N=0$.}
    \label{fig:img1}
  \end{subfigure}
  \hfill
  \begin{subfigure}{0.45\textwidth}
    \centering
    \includegraphics[width=\textwidth]{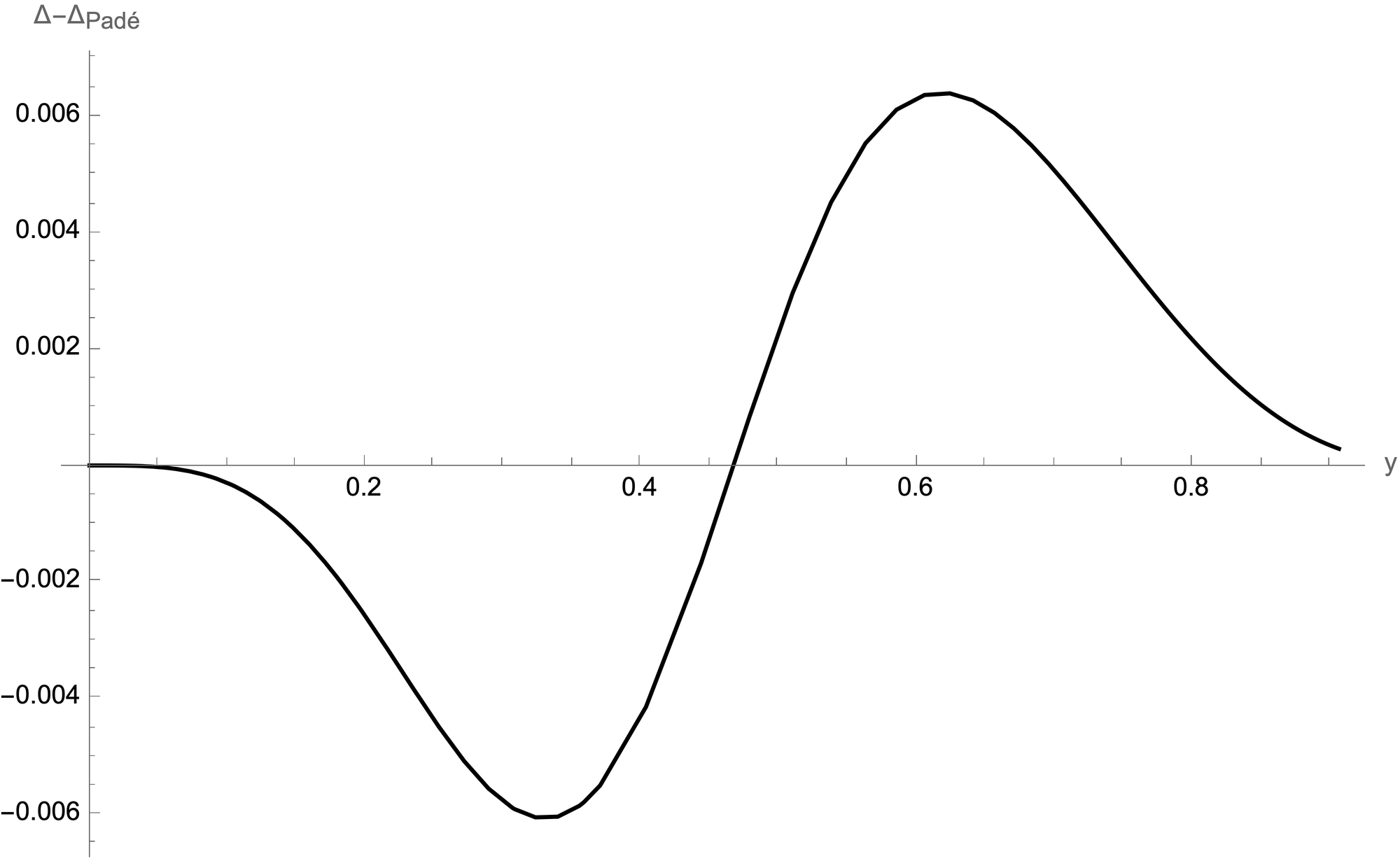}
    \caption{Error for $\Delta_{2,2}$ $M=3$ and $N=3$. }
    \label{fig:img2}
  \end{subfigure}
  \\[0.5cm]  
  \begin{subfigure}{0.45\textwidth}
    \centering
    \includegraphics[width=\textwidth]{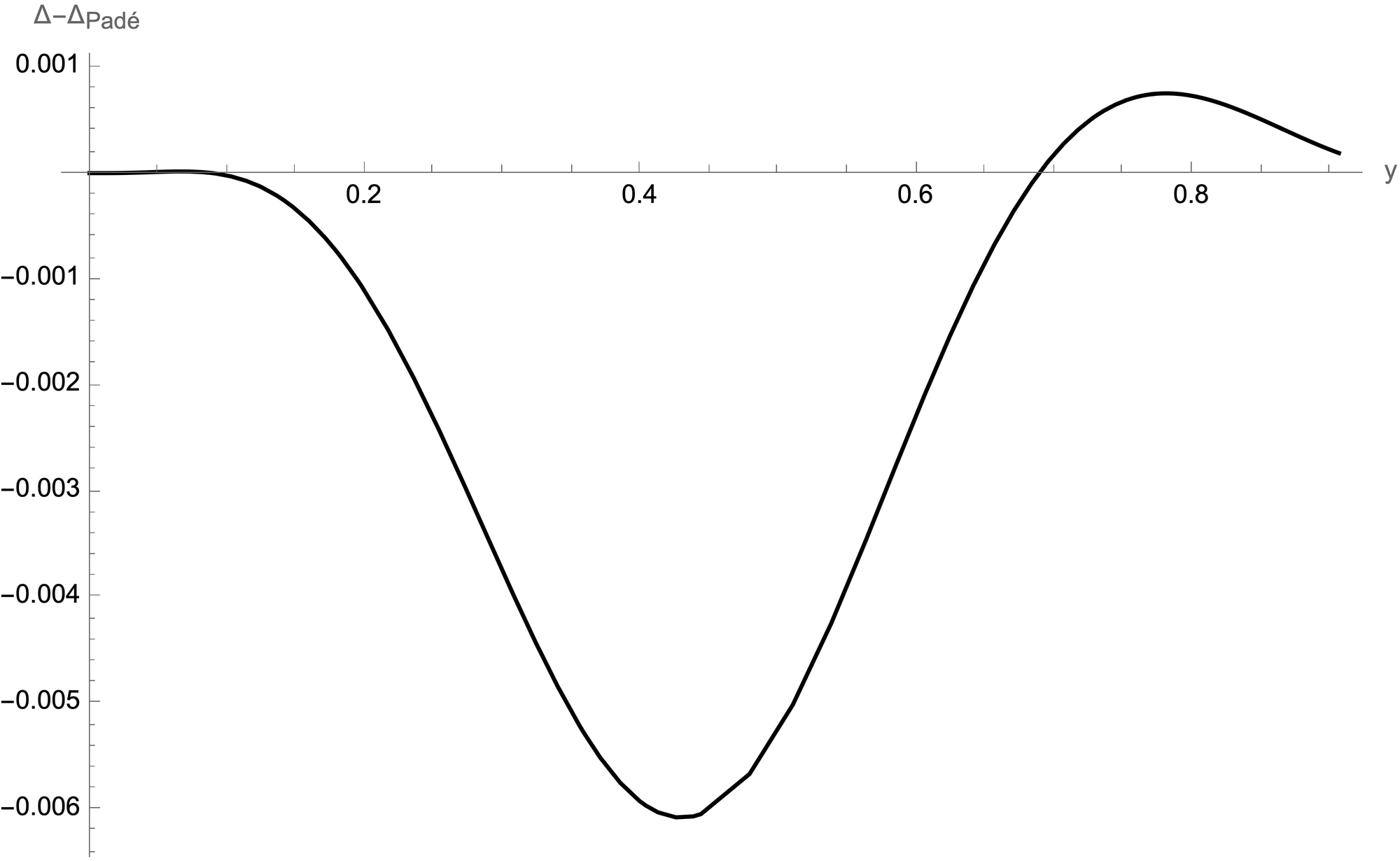}
    \caption{Error for $\Delta_{3,1}$ with $M=4$, $N=2$}.
    \label{fig:img3}
  \end{subfigure}
  \hfill
  \begin{subfigure}{0.45\textwidth}
    \centering
    \includegraphics[width=\textwidth]{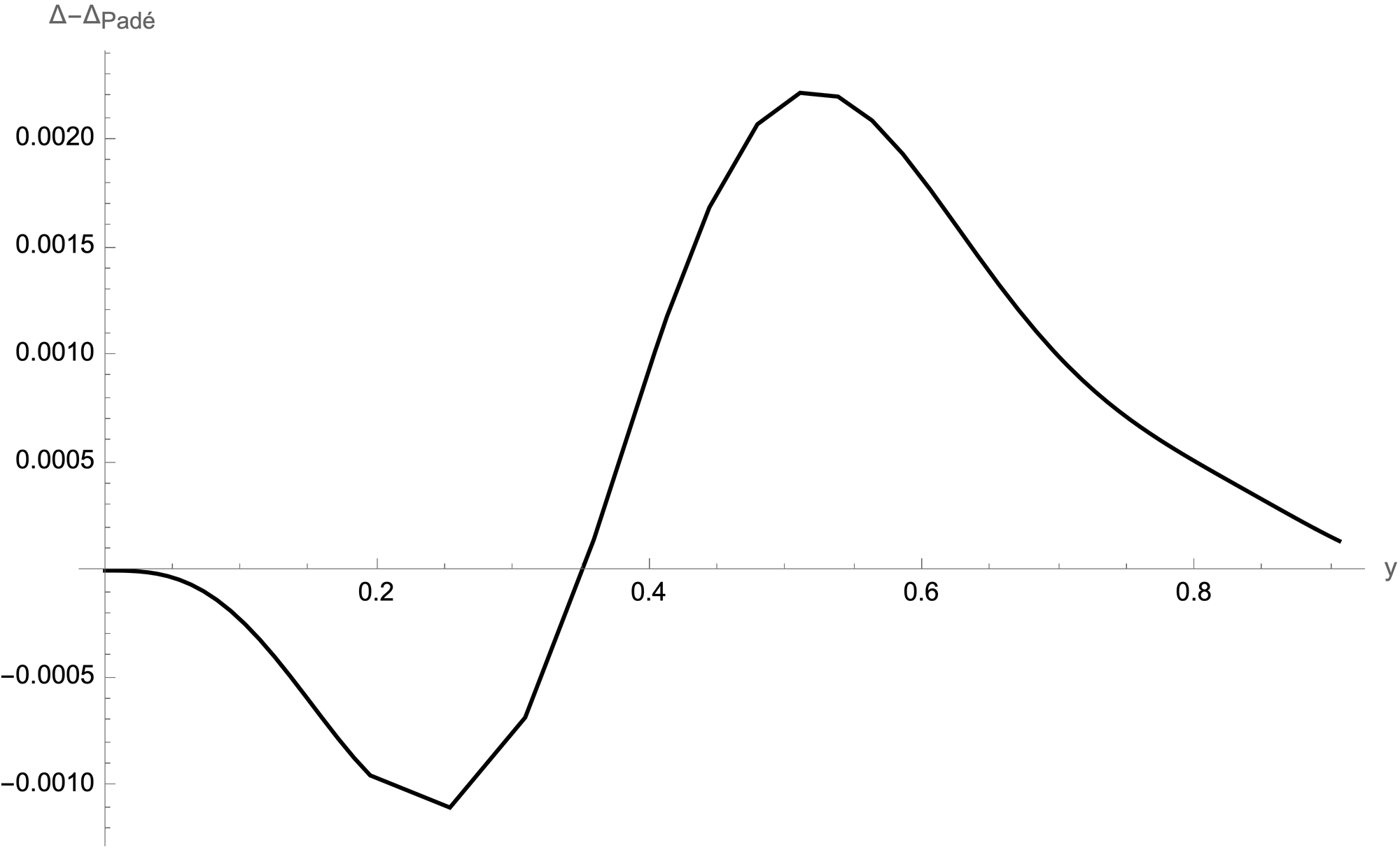}
    \caption{Error for $\Delta_{3,2}$ with $M=6$ and $N=0$.}
    \label{fig:img4}
  \end{subfigure}
  \caption{The difference between the exact spectrum from integrability and the best respective Pad\'e approximants for  $\Delta_{2,1}$, $\Delta_{2,2}$, $\Delta_{3,1}$ and $\Delta_{3,2}$.}
  \label{fig:PadeError}
  \end{figure}

\paragraph{Two sub-leading orders at strong coupling.}
Let us first concentrate on two-particle singlet states with $\Delta_0=2$ and $3$. There are two of each, which we denote respectively as $\Delta_{2,1}$, $\Delta_{2,2}$ and $\Delta_{3,1}$, $\Delta_{3,2}$.
The scaling dimension up to three sub-leading orders at weak coupling for these states~\cite{Cavaglia:2022qpg} can be obtained by solving the QSC analytically using the method described in~\cite{1DSolver}. At strong coupling, the scaling dimension is known up to two sub-leading orders from~\cite{Ferrero:2021bsb}.
Perturbatively, these states are expanded in the new coupling $y$, as
\begin{align}
    \label{eqn:PadePertWeak}
    \Delta_{\Delta_0,n}^\texttt{weak} &= \Delta_0 + a_{\Delta_0,n}^{(1)}\, y + a_{\Delta_0,n}^{(2)}\, y^2 + a_{\Delta_0,n}^{(3)}\, y^3 + \mathrm{O}(y^4)\;,  \\
    \label{eqn:PadePertStrong}
    \Delta_{\Delta_0,n}^\texttt{strong} &= 2\,\Delta_0 + b_{\Delta_0,n}^{(1)}\, (y-1) + b_{\Delta_0,n}^{(2)}\, (y-1)^2 + \mathrm{O}((y-1)^3)\;.
\end{align}
Thus, there are $7$ possible choices for Pad\'e approximants with the constraint $M+N=6$ with $M,N\geq0$. We summarises the choices that give the least error in Table~\ref{tab:2ptPadeBestDel23} below. We have
\begin{xltabular}[c]{\textwidth}{C|C|C|C}
\text{State} &
 M &
 N &
 \text{Max Error Order} 
 \\
 \midrule\midrule
 \rule{0pt}{3.5ex} 
 \Delta_{2,1} & 6 & 0 & 5\times 10^{-3}
 \\[1ex]
\hline
 \rule{0pt}{3.5ex} 
 \Delta_{2,2} & 3 & 3 & 6\times 10^{-3}
 \\[1ex]
\hline
 \rule{0pt}{3.5ex} 
 \Delta_{3,1} & 4\text{ or }5 & 2\text{ or }1 & 6\times 10^{-3}
 \\[1ex]
\hline
 \rule{0pt}{3.5ex} 
 \Delta_{3,2} & 6 & 0 & 2\times 10^{-3}
 \\[1ex]
\hline
  \caption{
  Two-sided Pad\'e approximants for states where we input three sub-leading orders at weak coupling and two sub-leading orders at strong coupling.
  }
  \label{tab:2ptPadeBestDel23}
\end{xltabular}
The results of the double-sided Pad\'e curves for these choices of $M$ and $N$, together with the integrability-based spectrum are shown on Figure~\ref{fig:Pade} with the difference between those curves shown explicitly on Figure~\ref{fig:PadeError}.

\paragraph{One sub-leading order at strong coupling.}
For general states in the 1D dSCFT, only one sub-leading order is known at strong coupling, due to~\cite{Ferrero:2021bsb}.
For the two-particle singlet states, it is relatively straightforward to generate up to three sub-leading orders, using the method described in~\cite{1DSolver}.
Given that we also know the non-perturbative scaling dimensions for the lowest lying of these states, 
one can attempt to benchmark the Pad\'e approximation on these states, before trying to apply it to states whose exact scaling dimensions are not known non-perturbatively. 
At weak coupling, the scaling dimensions are parameterised perturbatively exactly like equation~\eqref{eqn:PadePertWeak}. At strong coupling, they are parameterised as equation~\eqref{eqn:PadePertStrong}, however, with $b_{\Delta_0,n}^{(2)}=0$.
The Pad\'e Approximants have the same form~\eqref{eqn:PadeFunc}, however now the limits of the sums are subject to the constraint $M+N=5$. We find the least error for the following choices, given below in Table~\ref{tab:2ptPadeBestDel4567}. 
We study the three states with $\Delta_0 = 4,5$ and the four states with $\Delta_0 = 6,7$.
We find
\begin{xltabular}[c]{\textwidth}{C|C|C|C|C}
\text{State} &
 \Delta = \Delta_0 + \#\,g^2
 &
 M &
 N &
 \text{Max Error Order} 
 \\
 \midrule\midrule
 \rule{0pt}{3.5ex} 
 \Delta_{4,1} & 11.59133510 & 5  & 0 & 2\times 10^{-2}
 \\[1ex]
\hline
 \rule{0pt}{3.5ex} 
 \Delta_{4,2} & 8.359231442 & 3 & 2 & 8\times 10^{-3}
 \\[1ex]
\hline
 \rule{0pt}{3.5ex} 
 \Delta_{4,3} & 4.382766792 & 2 & 3 & 7\times 10^{-3}
 \\[1ex]
 \hline\hline
 \rule{0pt}{3.5ex} 
 \Delta_{5,1} & 13.11938368 & 3  & 2 & 6\times 10^{-2}
 \\[1ex]
\hline
 \rule{0pt}{3.5ex} 
 \Delta_{5,2} & 10.58150425 & 1 & 4 & 6\times 10^{-3}
 \\[1ex]
\hline
 \rule{0pt}{3.5ex} 
 \Delta_{5,3} & 6.765778741 & 3 & 2 & 2\times 10^{-2}
 \\[1ex]
\hline
\hline
 \rule{0pt}{3.5ex} 
 \Delta_{6,1} & 14.39810679 & 3  & 2 & 1\times 10^{-1}
 \\[1ex]
\hline
 \rule{0pt}{3.5ex} 
 \Delta_{6,2} & 12.34291807 & 2 & 3 & 2\times 10^{-2}
 \\[1ex]
\hline
 \rule{0pt}{3.5ex} 
 \Delta_{6,3} & 9.198904891 & 3 & 2 & 5\times 10^{-3}
 \\[1ex]
 \hline
 \rule{0pt}{3.5ex} 
 \Delta_{6,4} & 5.526736913 & 2 & 3 & 6\times 10^{-2}
 \\[1ex]
\hline
\hline
 \rule{0pt}{3.5ex} 
 \Delta_{7,1} & 15.49858743 & 3 & 2 & 1\times 10^{-1}
 \\[1ex]
\hline
 \rule{0pt}{3.5ex} 
 \Delta_{7,2} & 13.77362317 & 3 & 2 & 2\times 10^{-2}
 \\[1ex]
\hline
 \rule{0pt}{3.5ex} 
 \Delta_{7,3} & 11.27347285 & 2\text{ or }3 & 3\text{ or }2  & 2\times 10^{-2}
 \\[1ex]
 \hline
 \rule{0pt}{3.5ex} 
 \Delta_{7,4} & 7.625745134 & 2\text{ or }3 & 3\text{ or }2 & 2\times 10^{-2}
 \\[1ex]
\hline
  \caption{
  Two-sided Pad\'e approximants for states where we input three sub-leading orders at weak coupling and one sub-leading order at strong coupling.
  We also display the one-loop anomalous dimensions of these states at weak coupling in the second column. We observe that the magnitude of one-loop anomalous dimension correlates with the maximal error in the Pad\'e approximants. 
  }
  \label{tab:2ptPadeBestDel4567}
\end{xltabular}
There are three observations that we can draw from the results above:
\begin{itemize}
    \item It seems that the closest approximations are obtained for those states whose one-loop anomalous dimension is around halfway between the maximal and minimal allowed value for all the states with that value of $\Delta_0$. 
    \item It also seems that the preferred values of $M$ and $N$ are those that give a more ``square'' rather than ``rectangular'' approximant, meaning the numerator and denominator in the rational function~\eqref{eqn:PadeFunc} tend to have similar degrees. There are of course exceptions to this, especially in the case where more orders are known at strong coupling. 
    \item The maximal error seems to get worse as $\Delta_0$ increases.
\end{itemize}
We also note that the sub-optimal Pad\'e approximants display various kinds of behaviours: some of them develop poles in $y\in[0,1]$, 
some of them give a much higher maximal error than the optimal Pad\'e approximant, whereas sometimes the sub-optimal Pad\'e approximants are pretty close to the optimal ones.
We have not been able to quantify which parameter(s), if any, control(s) this variance in the behaviour of the sub-optimal Pad\'e approximants. 
\paragraph{Thumb Rules.}
Based on these conclusions, we propose the following thumb rules for performing a Pad\'e approximation for a state not just in the 1D dSCFT, but also in related setups, for instance, the one studied in~\cite{Gabai:2025hwf,FluxTubePade}, where there is some perturbative data available at both weak and strong coupling, but no non-perturbative data at finite coupling:
\begin{enumerate}
    \item If there is a multiplicity of states, then the state whose anomalous dimension is in the middle of the range of possible anomalous dimensions should yield most accurate results. 
    \item Prefer square to rectangular Pad\'e approximants.
    \item There may still be significant deviation from the exact value at finite coupling, but the data may be useful for bootstrap applications at small/large but finite coupling, outside the purview of perturbation theory.
\end{enumerate}

It is also possible to do a one-sided Pad\'e approximation. This turns out to be much less accurate than the two-sided Pad\'e approximation considered above. We present our results about one-sided Pad\'e approximations in Appendix~\ref{adx:Pade}.

%% file: disc_v2.tex
\section{Discussion}
\label{sec:discussion}

In this paper, we have discussed the possibility of providing a smooth map between weak (gauge) and strong (string) coupling degrees of freedom for the supersymmetric Wilson line-defect in planar $\mathcal{N} = 4$ SYM. 
By analysing the Partition Functions computed at zero and infinite coupling, we proposed a map from 
degrees of freedom of the single-particle and singlet two-particle sectors of the theory at weak coupling, 
to strong coupling.
The map predicted 
that the scaling dimension of these states doubles from zero to infinite coupling.
We tested this prediction 
by comparing with
the non-perturbative spectrum obtained using the QSC, and found agreement.

It was crucial that the two-particle words form super-primary states among themselves, at one-loop at weak coupling, in order to generate and test a prediction for the behaviour of their scaling dimensions. 
However, the authors of~\cite{Ferrero:2021bsb,Ferrero:2023gnu} observed that the eigenstates of the Dilatation operator at strong coupling \emph{do} mix two- and higher-particle words.
That our map still persists
could suggest 
a potential hidden symmetry in the supersymmetric Wilson line-defect in $\mathcal{N} = 4$ SYM, that allows us to further grade super-primary states by some effective particle number that is related to their particle-number at one-loop.
Combined with the fact that integrability allows level-crossings, 
this could also explain 
how our prediction 
$\Delta_\infty = 2\Delta_0$
could be consistent with the generic expectation
that a smooth map of degrees of freedom would imply $\Delta_0\to \Delta_\infty = \Delta_0^2$~\cite{Gabai:2025hwf,FluxTubePade}.

It would also be extremely interesting to make contact with recent developments in full $\mathcal{N} = 4 $ SYM~\cite{Gopakumar:2024jfq}, where 
the gauge theory Feynman diagram expansion was recast as a sum over dual closed strings.

Let us discuss potential next steps forward. 
First of all, it would be beneficial to complete the full QSC description of insertions on the supersymmetric Wilson line, by using the full prescription~\eqref{eqn:map1D4D} from the 1D dSCFT to $\mathcal{N} = 4$ SYM~\cite{1DSolver}.
This would help to propose and test further maps of degrees of freedom in other sectors of the theory. 
Knowledge of the spectrum of states with transverse spin about the Wilson line could also be useful in the context of recent $S$-matrix bootstrap calculations~\cite{Alday:2025pmg}.
It would be also be very insightful to have a full one-loop spin-chain Hamiltonian capturing insertions on the supersymmetric Wilson line.
This would enable the explicit construction of eigenstates to further probe the structure of eigenstates~\cite{Full1DHam}. 

There are potential practical implementations for the map that we propose. 
If at least some perturbative orders are available for more highly excited two-particle states with large bare dimensions $\Delta_0$, 
then this map would enable a double-sided
Pad\'e approximation, 
to generate predictions for the scaling dimensions at intermediate coupling without much effort.
This spectral data could be useful input to bootstrap observables in theory at strong coupling
\cite{Ghosh:2025sic,Funbootstrability1}, mirroring similar developments in $\mathcal{N}=4$ SYM~\cite{Julius:2023hre,Julius:2024ewf}.

One can of course, improve the formulation of the map itself: to make it more precise, and to propose a mapping between characters at weak and storing coupling similar to what was done in the $\mathcal{N} = 4$ SYM context~\cite{Spradlin:2004pp}.
Analysis of characters have also proven useful in the analysis and classification of of 2D CFTs~\cite{Mathur:1988na,Downing:2025huv}.
Another interesting direction is to calculate the super-conformal index -- and to see if the index gives more insights between mapping weak and strong degrees of freedom.

For this case, we touched only on the supersymmetric Wilson line defect in the planar, large-$N$, limit. It would be interesting to go away from the large-$N$, evaluate Partition Functions and to see if the proposed map persists. The same procedure can be in principle repeated for a non-supersymmetric Wilson line-defects, or for setups with cusps.

In conclusion, it really does seem that the supersymmetric Wilson line-defect in $\mathcal{N}=4$ SYM, functions as a Rosetta Stone for Wilson-line defects, and we hope that conclusions drawn from studying it would be helpful in understanding 
the map between gauge and string of degrees of freedom.

%% file: supermultiplets.tex
\section{Supermultiplets}
\label{app:supermultiplets}

The details of the superconformal algebra were given in~\cite{Agmon:2020pde} and the construction of the supermultiplets in the 1D dSCFT was performed in~\cite{Ferrero:2023znz}. Below we display the display the relevant results from these papers for convenience of the reader.

The supercharges transform under the fundamental representation of $\mathrm{SU}(2)\times \mathrm{Sp}(4)$:
\begin{equation}
    \mathcal{Q}_{a\,\alpha} \in [1]_{1/2}^{(1, 0)},\ \ \ \text{with }\ \\ a=1,\dots,4\;,\quad \alpha = 1,2\;.
\end{equation}
where weights can be assigned to each individual supercharge as shown in Table \ref{tab:supercharges}.
\begin{table}[h!]
\centering
\renewcommand{\arraystretch}{1}
\setlength{\tabcolsep}{8pt}
\setlength{\extrarowheight}{4pt}
\begin{tabular}{ |c|c|c|c| } 
 \hline
 $\mathcal{Q}_{11}[+1]_{1/2}^{(1, 0)}$ & $\mathcal{Q}_{21}[+1]_{1/2}^{(-1, 1)}$ & $\mathcal{Q}_{31}[+1]_{1/2}^{(1, -1)}$ & $\mathcal{Q}_{41}[+1]_{1/2}^{(-1, 0)}$  \\ 
 \hline
 $\mathcal{Q}_{12}[-1]_{1/2}^{(1, 0)}$ & $\mathcal{Q}_{22}[-1]_{1/2}^{(-1, 1)}$ & $\mathcal{Q}_{32}[-1]_{1/2}^{(1, -1)}$ & $\mathcal{Q}_{42}[-1]_{1/2}^{(-1, 0)}$  \\ 
 \hline
\end{tabular}
\caption{Charges of the super-charges~\cite{Agmon:2020pde}.} 
\label{tab:supercharges}
\end{table}
The detailed description of super-multiplets is shown in Table~\ref{tab:multiplets}.
\renewcommand{\arraystretch}{1} 
\setlength{\extrarowheight}{3pt} 
\newcolumntype{P}[1]{>{\centering\arraybackslash}p{#1}}
\begin{table}[h!]
\centering
\begin{tabular}{ |P{1cm}|P{3cm}|P{3.7cm}|P{1cm}|P{3cm}| } 
 \hline
 \textbf{Type} & \textbf{Super-primary} & \textbf{Unitarity Bound} & \textbf{BPS} & \textbf{Null State}  \\ 
 \hline
 $\mathcal{L}$ & $\displaystyle [j]_{\Delta}^{(R_1,R_2)}$ & $\displaystyle \Delta > \tfrac{1}{2}j + R_1 + R_2 + 1$ & $-$ & $-$  \\ 
 \hline
 $\mathcal{A}_1$ & $\displaystyle [j]_{\Delta}^{(R_1,R_2)}\ (j>0)$ & $\displaystyle \Delta = \tfrac{1}{2}j + R_1 + R_2 + 1$ & $\tfrac{1}{8}$ & $\displaystyle [j-1]_{\Delta+1/2}^{(R_1+1,R_2)}$\\
 \hline
  & $\displaystyle [j]_{\Delta}^{(0,R_2)}$ & $\displaystyle \Delta = \tfrac{1}{2}j + R_2 + 1$ & $\tfrac{1}{4}$ & $\displaystyle [j-1]_{\Delta+1/2}^{(1,R_2)}$ \\ 
 \hline
  & $\displaystyle [j]_{\Delta}^{(0,0)}$ & $\displaystyle \Delta = \tfrac{1}{2}j + 1$ & $\tfrac{1}{2}$ & $\displaystyle [j-1]_{\Delta+1/2}^{(1,0)}$  \\ 
 \hline
 $\mathcal{A}_2$ & $\displaystyle [0]_{\Delta}^{(R_1,R_2)}$ & $\displaystyle \Delta = R_1 + R_2 + 1$ & $\tfrac{1}{8}$ & $\displaystyle [0]_{\Delta+1}^{(R_1+2,R_2)}$ \\ 
 \hline
  & $\displaystyle [0]_{\Delta}^{(0,R_2)}$ & $\displaystyle \Delta = R_2 + 1$ & $\tfrac{1}{4}$ & $\displaystyle [0]_{\Delta+1}^{(2,R_2)}$ \\ 
 \hline
  & $\displaystyle [0]_{\Delta}^{(0,0)}$ & $\displaystyle \Delta = 1$ & $\tfrac{1}{2}$ & $\displaystyle [0]_{\Delta+1}^{(2,0)}$  \\ 
 \hline
 $\mathcal{B}_1$ & $\displaystyle [0]_{\Delta}^{(R_1,R_2)}$ & $\displaystyle \Delta = R_1 + R_2$ & $\tfrac{1}{4}$ & $\displaystyle [1]_{\Delta+1/2}^{(R_1+1,R_2)}$  \\ 
 \hline
 $\mathcal{B}_1$ & $\displaystyle [0]_{\Delta}^{(0,R_2)}$ & $\displaystyle \Delta = R_2$ & $\tfrac{1}{2}$ & $\displaystyle [1]_{\Delta+1/2}^{(1,R_2)}$  \\ 
 \hline
\end{tabular}
\caption{Superconformal multiplets and thier classification, which were worked out in \cite{Agmon:2020pde}. The notation $\mathcal{L}$-type , $\mathcal{A}$-type and $\mathcal{B}$-type stand for long, semi-short and short super-multiplets respectively.}
\label{tab:multiplets}
\end{table}

The explicit content of the 
most important super-multiplets, worked out in~\cite{Ferrero:2023znz}, are displayed below 
\begin{equation}
\mathcal{B}_{1}: [0]^{(0, 1)}_{1} \to [1]^{(1, 0)}_{3/2} \to [2]^{(0, 0)}_{2}
\end{equation}
\begin{gather}
    \mathcal{A}_{1} : [2]^{(0, 0)}_{2}  \to [3]^{(1, 0)}_{\frac{5}{2}}  \to [4]^{(0, 0)}_{3} \oplus [4]^{(0, 1)}_{3} \to [5]^{(1, 0)}_{\frac{7}{2}} \to [6]^{(0, 0)}_{4}
\end{gather}
\begin{gather}
    \mathcal{A}_{2} : [0]^{(0, 0)}_{1}  \to [1]^{(1, 0)}_{\frac{3}{2}}  \to [2]^{(0, 0)}_{2} \oplus [2]^{(0, 1)}_{2} \to [3]^{(1, 0)}_{\frac{5}{2}} \to [4]^{(0, 0)}_{3}
\end{gather}
\begin{gather}
    \mathcal{L} : [0]^{(0, 0)}_{2} \to \nonumber \\ \to [1]^{(1, 0)}_{\frac{5}{2}} \to \nonumber \\  \to [2]^{(0, 0)}_{3} \oplus [2]^{(0, 1)}_{3} \oplus [0]^{(2, 0)}_{3} \to \nonumber \\  \to [1]^{(1, 0)}_{\frac{7}{2}} \oplus [3]^{(1, 0)}_{\frac{7}{2}} \oplus [1]^{(1, 1)}_{\frac{7}{2}} \to \nonumber \\  \to [1]^{(0, 0)}_{0} \oplus [1]^{(0, 0)}_{4} \oplus [1]^{(0, 1)}_{4} \oplus [1]^{(0, 1)}_{4} \oplus [1]^{(0, 2)}_{4} \oplus [1]^{(2, 0)}_{4} \to \nonumber \\ \to [1]^{(1, 0)}_{\frac{9}{2}} \oplus [3]^{(1, 0)}_{\frac{9}{2}} \oplus [1]^{(1, 1)}_{\frac{9}{2}} \to \nonumber \\ \to [2]^{(0, 0)}_{5} \oplus [2]^{(0, 1)}_{5} \oplus [0]^{(2, 0)}_{5} \to \nonumber  \\  \to [1]^{(1, 0)}_{\frac{11}{2}} \to \nonumber \\ \to [0]^{(0, 0)}_{6}
\end{gather}

%% file: pade_adx.tex
\section{One-sided Pad\'e Approximation}\label{adx:Pade}
\begin{figure}[H]
  \centering
  \begin{subfigure}{0.45\textwidth}
    \centering
    \includegraphics[width=\textwidth]{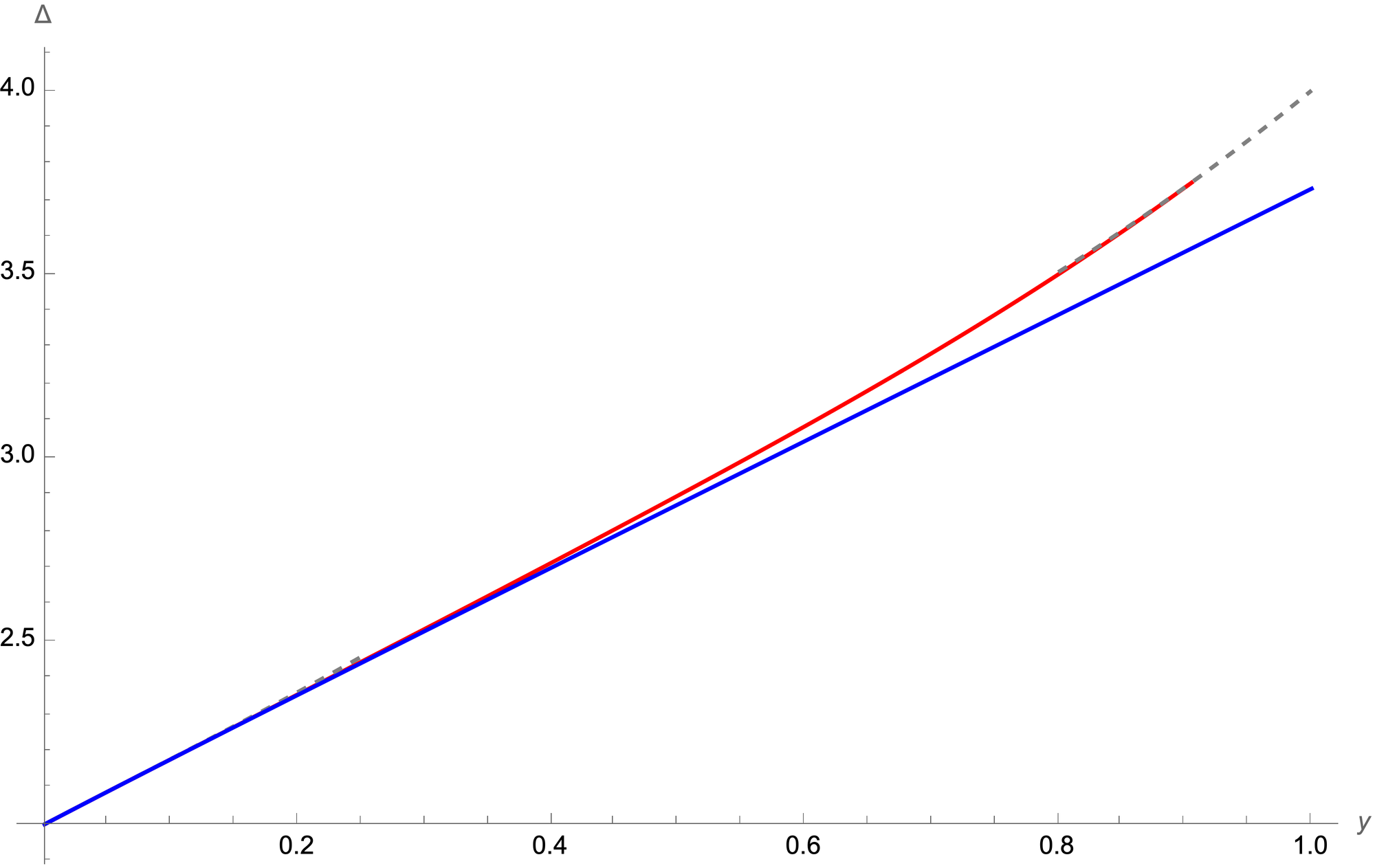}
    \caption{One-sided Pad\'e approximation for $\Delta_{2,1}$ inputting three sub-leading orders at weak coupling.}
    \label{fig:img1}
  \end{subfigure}
  \hfill
  \begin{subfigure}{0.45\textwidth}
    \centering
    \includegraphics[width=\textwidth]{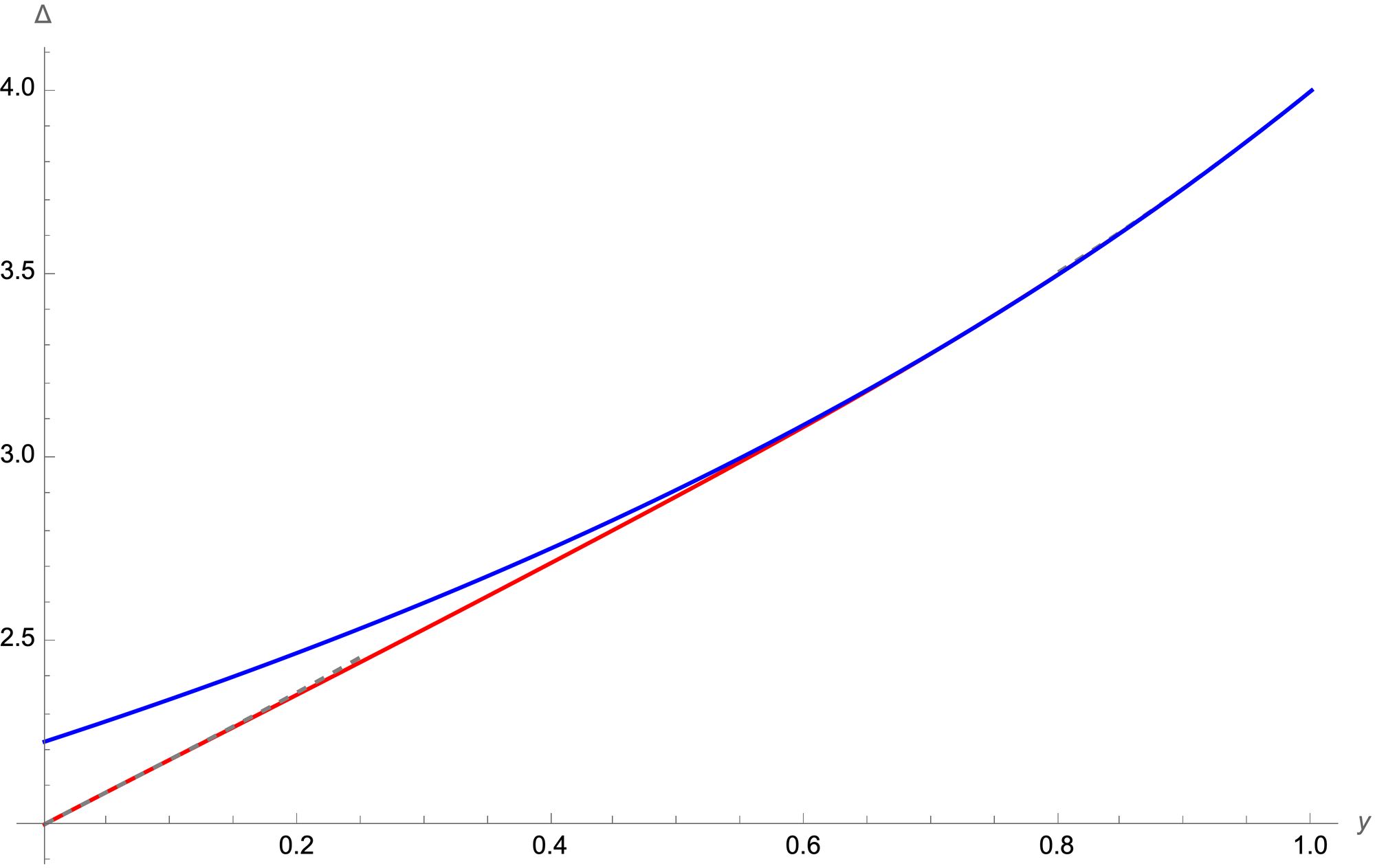}
    \caption{One-sided Pad\'e approximation for $\Delta_{2,1}$ inputting two sub-leading orders at strong coupling.}
    \label{fig:img2}
  \end{subfigure}
  \\[0.5cm]  
  \begin{subfigure}{0.45\textwidth}
    \centering
    \includegraphics[width=\textwidth]{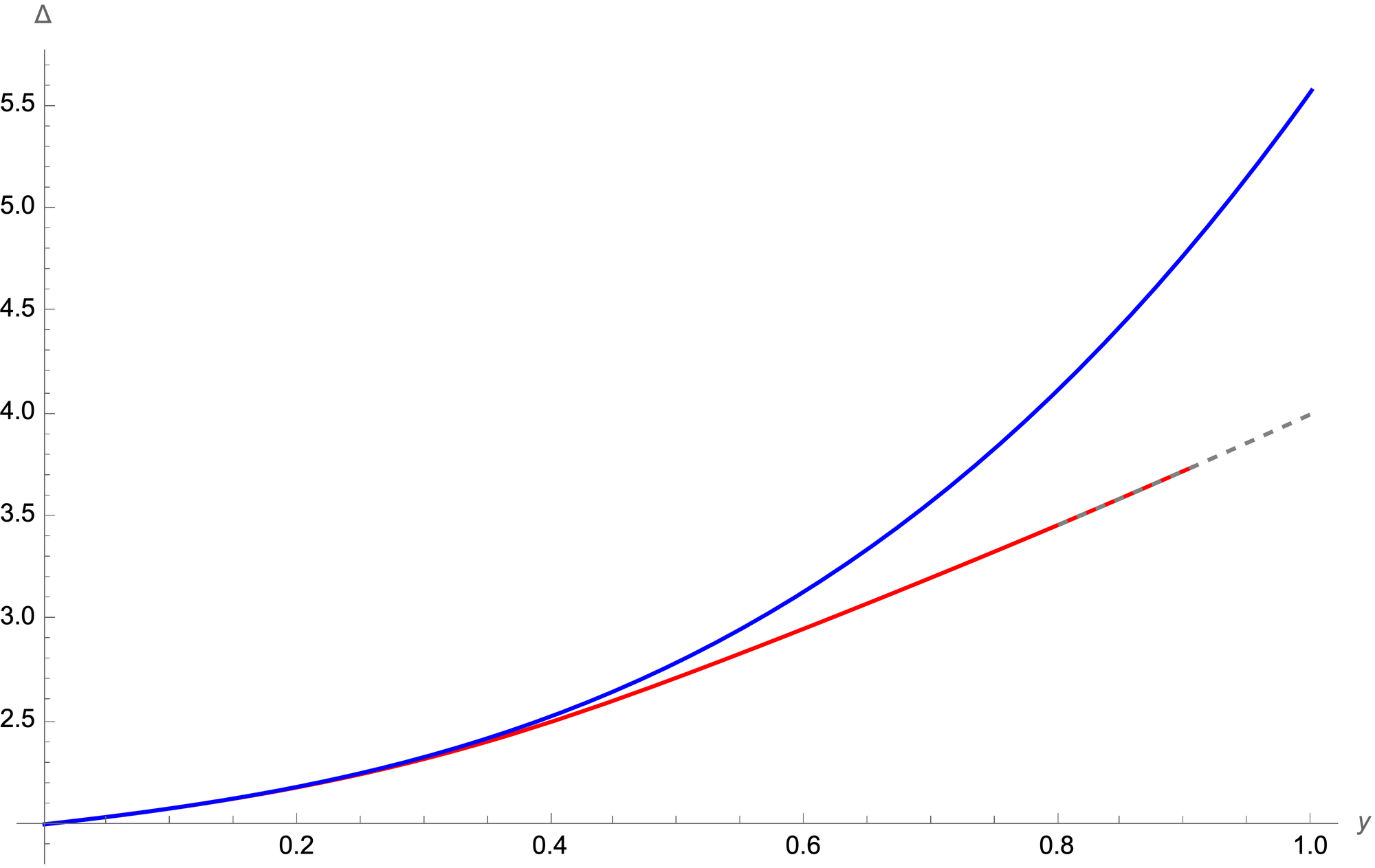}
    \caption{One-sided Pad\'e approximation for $\Delta_{2,2}$ inputting three sub-leading orders at weak coupling.}
    \label{fig:img3}
  \end{subfigure}
  \hfill
  \begin{subfigure}{0.45\textwidth}
    \centering
    \includegraphics[width=\textwidth]{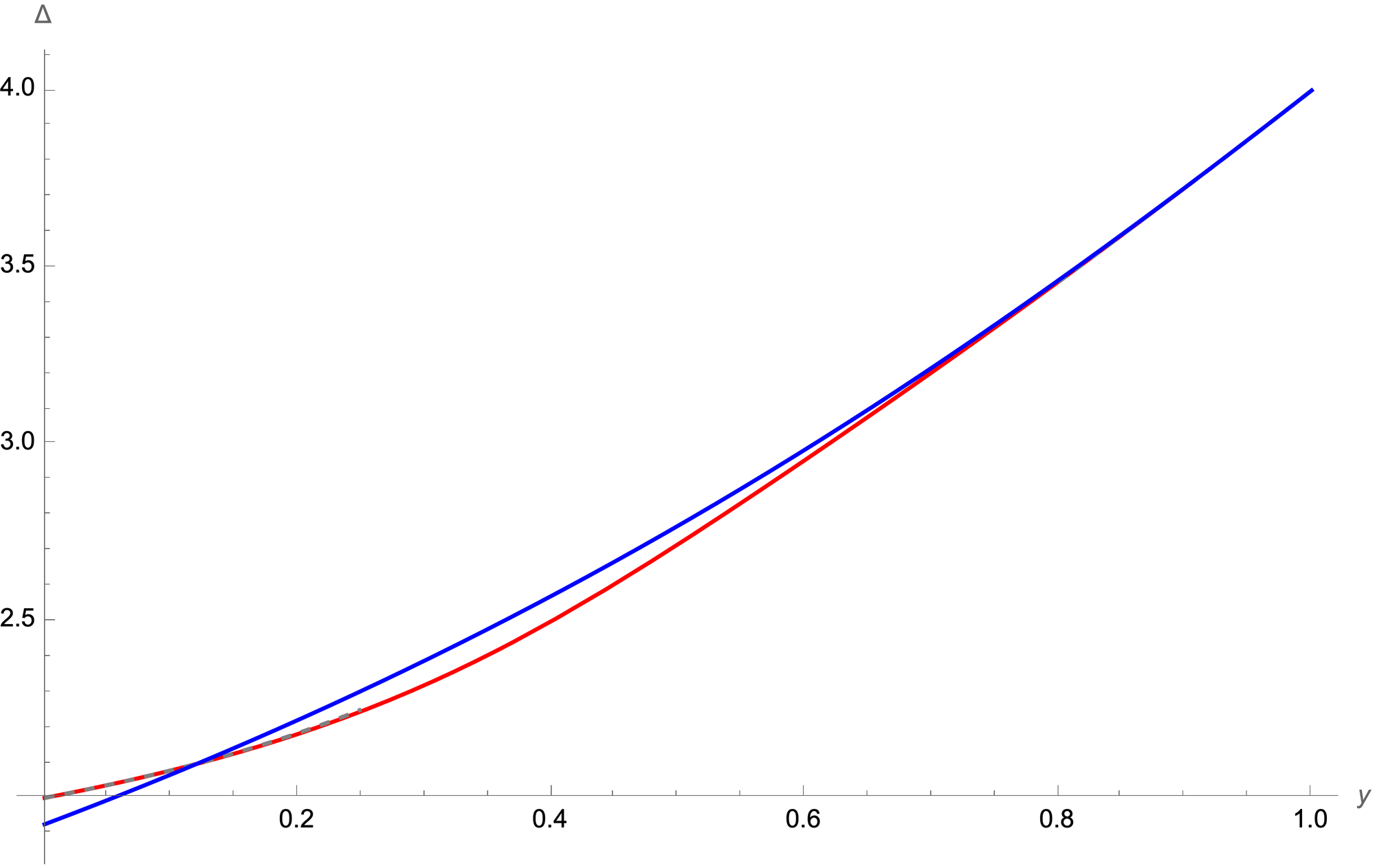}
    \caption{One-sided Pad\'e approximation for $\Delta_{2,1}$ inputting two sub-leading orders at strong coupling.}
    \label{fig:img4}
  \end{subfigure}
\caption{The best one-sided Pad\'e approximants for $\Delta_{2,1}$ and $\Delta_{2,2}$.
The exact integrability data is shown in red, and the approximation in blue. The weak and strong perturbative data is shown with the dashed lines.
}
  \label{fig:PadeSingle}
\end{figure}

We display our results for the best one-sided Pad\'e approximants for $\Delta_{2,1}$ and $\Delta_{2,2}$ in Figure~\ref{fig:PadeSingle}.
